\documentclass[acmtog, balance = false]{acmart}
\acmJournal{TOG}
\setcitestyle{square}

\usepackage{wrapfig}
\usepackage{nicefrac}
\usepackage{mathtools}
\usepackage{graphicx}
\usepackage{booktabs} 
\usepackage{combelow} % for comma underneath a word
\usepackage[ruled,vlined,linesnumbered]{algorithm2e}
\usepackage{tabularx} % beautiful table
\usepackage{colortbl} % add row color to a table

\usepackage{manfnt} % cube symbol

\usepackage{amsmath}
\usepackage{amsfonts}
\usepackage{amssymb}
\usepackage{amsbsy}

\usepackage[mathletters]{ucs}
\usepackage[utf8x]{inputenc}

\definecolor{derekBlue}{RGB}{144,210,236}
\definecolor{derekTableBlue}{RGB}{189,235,252}
\definecolor{iglGreen}{RGB}{153,203,67}
\definecolor{coralRed}{RGB}{250,114,104}
\definecolor{gray}{RGB}{180,180,180}

\SetCommentSty{commentFont}
\SetNlSty{textbf}{}{.}

\newcommand{\update}[1]{#1}

\newcommand{\reffig}[1] {Fig.~\ref{fig:#1}}

\newcommand{\reftab}[1] {Table~\ref{tab:#1}}
\newcommand{\refsec}[1] {Sec.~\ref{sec:#1}}
\newcommand{\refapp}[1] {App.~\ref{app:#1}}

\newcommand{\R}{\mathbb{R}}
\newcommand{\M}{\mathcal{M}}
\newcommand{\N}{\mathcal{N}}

\newcommand{\MAPS}{\textsc{MAPS}\ }
\newcommand{\qslim}{\textsc{qslim}\ }
\newcommand{\vertex}{\textsc{Vertex}\ }
\newcommand{\edge}{\textsc{Edge}\ }
\newcommand{\initialization}{\textsc{Initialization}\ }
\newcommand{\V}{$\mathcal{V}$\ }
\newcommand{\E}{$\mathcal{E}$\ }
\newcommand{\I}{$\mathcal{I}$\ }

\newcommand{\vecFont}[1]{\mathsf{#1}}

\def\vb{{\vecFont{b}}}

\def\vn{\hat{\vecFont{n}}}

\def\vv{{\vecFont{v}}}

\newcommand{\matFont}[1]{\mathsf{#1}}

\def\mF{{\matFont{F}}}

\def\mV{{\matFont{V}}}

\SetAlFnt{\small}
\SetAlCapFnt{\small}
\SetAlCapNameFnt{\small}
\SetAlCapHSkip{0pt}

\acmJournal{TOG}
\acmYear{2020}\acmVolume{39}\acmNumber{4}\acmArticle{1}\acmMonth{7} \acmDOI{10.1145/3386569.3392418}

\begin{document}

\title{Neural Subdivision}

\author{Hsueh-Ti Derek Liu}
\affiliation{\institution{University of Toronto}}
\email{hsuehtil@cs.toronto.edu}

\author{Vladimir G. Kim}
\affiliation{\institution{Adobe Research}}
\email{vokim@adobe.com}

\author{Siddhartha Chaudhuri}
\affiliation{\institution{Adobe Research, IIT Bombay}}
\email{sidch@cse.iitb.ac.in}

\author{Noam Aigerman}
\affiliation{\institution{Adobe Research}}
\email{aigerman@adobe.com}

\author{Alec Jacobson}
\affiliation{\institution{University of Toronto, Adobe Research}}
\email{jacobson@cs.toronto.edu}

% A "teaser" figure, centered below the title and authors and above the body of the work.
\begin{teaserfigure}
  \includegraphics[width=\linewidth]{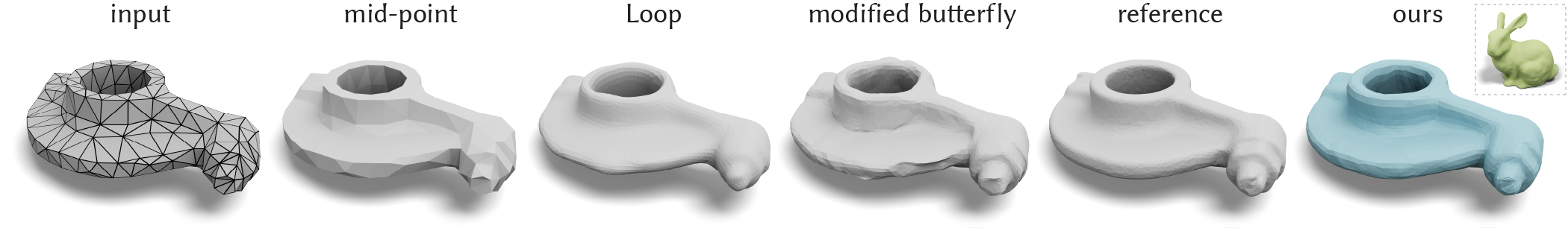}
  \caption{Our neural subdivision framework performs geometry-aware subdivision, reconstructing the reference rocker arm that we decimated to obtain the coarse input with high accuracy, even though it was only trained on one single model - the Stanford bunny. Neural subdivision does not suffer from the inherent limitations of classic subdivisions, such as volume shrinkage and over-smoothing (Loop~\shortcite{loop1987smooth}), or amplification of tessellation artifacts (modified butterfly~\cite{zorin1996interpolating}). Throughout this paper, we use green to denote the training shape, and blue for the neural subdivision output.}
  \label{fig:teaser}
\end{teaserfigure}
% \textsc{AIM@SHAPE} \cite{AIMSHAPE}

\begin{abstract}
This paper introduces \emph{Neural Subdivision}, a novel framework for data-driven coarse-to-fine geometry modeling. During inference, our method takes a coarse triangle mesh as input and recursively subdivides it to a finer geometry by applying the fixed topological updates of Loop Subdivision, but predicting vertex positions using a neural network conditioned on the local geometry of a patch. This approach enables us to learn complex non-linear subdivision \update{schemes}, beyond simple linear averaging used in classical techniques.
One of our key contributions is a novel self-supervised training setup that only requires a set of high-resolution meshes for learning network weights. For any training shape, we stochastically generate diverse low-resolution discretizations of coarse counterparts, while maintaining a bijective mapping that prescribes the exact target position of every new vertex during the subdivision process. This leads to a very efficient and accurate loss function for conditional mesh generation, and enables us to train a method that generalizes across discretizations and favors preserving the manifold structure of the output.
During training we optimize for the same set of network weights across all local mesh patches, thus providing an architecture that is not constrained to a specific input mesh, fixed genus, or category. Our network encodes patch geometry in a local frame in a rotation- and translation-invariant manner. Jointly, these design choices enable our method to generalize well, and we demonstrate that even when trained on a single high-resolution mesh our method generates reasonable subdivisions for novel shapes.

%This paper addresses a new problem on how to subdivide a 3D shape using neural networks, called \emph{neural subdivision}. Given a coarse mesh as input, our goal is to subdivide the mesh using a geometry-dependent rule learned by a neural network.
%
%The key challenge is to output a mesh that has both correct connectivity and vertex information, unlike traditional network upsampling tasks which solely focus on the vertex information.
%
%The lack of proper loss function that is aware of the manifold structure and means of creating training data prevent one from training subdivision networks.
%
%In response, we complete the neural subdivision framework by proposing a method for generating self-supervised training data given only high-resolution shapes, and defining the loss function that is aware of the mesh connectivity.
%
%The main building block is a two-step parameterization module that can be plugged into any edge decimation algorithm to construct a bijective map between the input mesh and the decimated mesh on-the-fly.
%
%In addition, we exploit the property of mesh structure and propose a subdivision network architecture that is invariant to rigid motions, can generalize to different input triangulations, and can subdivide unseen shapes even when trained on only one single shape.
%
%Our first steps towards neural subdivision could enable future work on developing data-driven subdivision methods for create meshes with different level of details.
\end{abstract}

% Processes all of the front-end information and starts the body of the work.
\maketitle
\section{Introduction}
\begin{figure}
    \centering
    \includegraphics[width=3.33in]{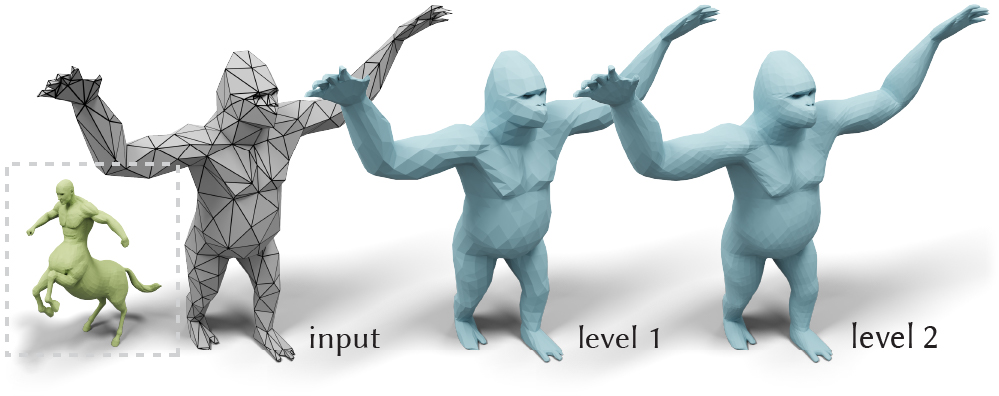}
    \vspace{-5pt}
    \caption{\update{Neural subdivision refines different parts of a mesh differently, conditioned on the local geometry. Here, the network was trained on the centaur model (green) and then evaluated on a coarse gorilla mesh (gray).} }
    \label{fig:allLevels}
    \vspace{-10pt}
\end{figure}
Subdivision surfaces are defined by deterministic, recursive upsampling of a
discrete surface mesh.
Classic methods work by performing two steps: each input mesh element is
\emph{divided} into many elements (e.g., one triangle becomes three) by
splitting edges and adding vertices.
The positions of the mesh vertices are then smoothed by taking a weighted
average of their neighbors' positions according to a weighting scheme based
purely on the local mesh connectivity.
Subdivision surfaces are well studied and have rich theory connecting their
limit surfaces (applying an infinite number of subdivide-and-smooth iterations)
to traditional splines.
They are a standard paradigm in surface modeling tools, allowing modelers to
sculpt shapes in a coarse-to-fine manner.
A modeler may start with a very coarse cage, adjust vertex positions, then
subdivide once, adjust the finer mesh vertices, and repeat this process until
satisfied.

While existing subdivision methods are well-suited for this sort of interactive
modeling, they fall short when used to automatically upsample a low resolution
asset.
Without a user's guidance, classic methods will overly smooth the entire shape
(see \reffig{teaser}).
Popular methods based on simple linear averaging do not identify details to
maintain or accentuate during upsampling.
They make no use of the geometric context of a local patch of a surface.
Furthermore, classic methods based on fixed one-size-fits-all weighting
rules are determined for their general convergence and smoothness properties.
This ignores an opportunity to leverage the massive amount of information
lurking in the wealth of existing 3D models.

We propose \emph{Neural Subdivision}.
We recursively subdivide an input triangle mesh by applying the same fixed
topological updates of classic Loop Subdivision, but move vertices according to
a neural network conditioned on the local patch geometry.
We train the shared weights of this network to learn a geometry-dependent
non-linear \update{subdivision} that goes beyond classic linear averaging (see \reffig{allLevels}).
The choice of training data tailors the network to a particular class, type or
diversity of geometries.

An immediate challenge is how to collect training data pairs.
There is an ever-growing number of 3D models available.
However, many if not most of them were not created using a subdivision modeling
tool. Even among those that were, the final model does not retain information
to replay the modeler's vertex displacements.
In the absence of paired data for a supervised training approach, we propose a
novel method to \emph{self-supervise} given only high-resolution surface meshes
of arbitrary origin/connectivity at training time.
We stochastically generate candidate low-resolution versions of a training
exemplar while maintaining a bijective correspondence between their surfaces.
This correspondence enables a novel loss function that is more efficient and
accurate compared to existing methods.
By construction, this training regime ensures \emph{generalization across
discretization}.

\begin{figure}
    \centering
    \includegraphics[width=3.33in]{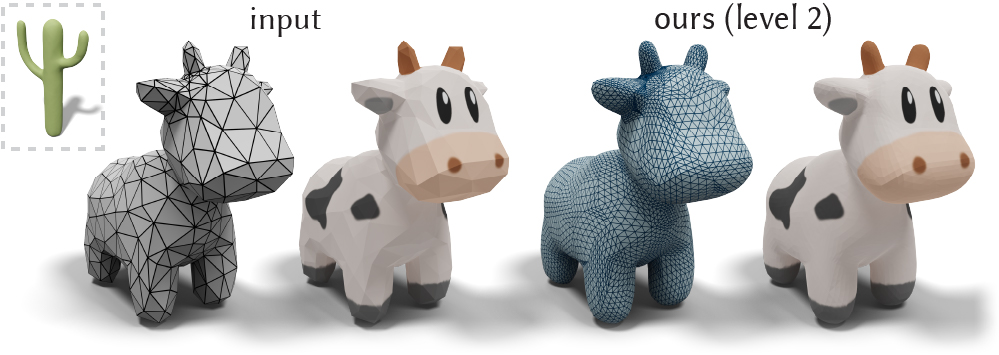}
    \vspace{-5pt}
    \caption{Neural subdivision can adapt to different input triangulations and output a high-resolution surface mesh accordingly. This enables us to use it directly in the graphics pipeline such as texture mapping.}
    \label{fig:texture}
    \vspace{-10pt}
\end{figure}
In contrast to existing generative models for surfaces, our output is a
surface mesh with deterministic connectivity based on the input, enabling direct
use in the standard graphics pipeline such as texture mapping (see \reffig{texture}).
By sharing weights and training across all local patches of all the training
meshes, we learn a rule based on the local neighborhood rather than the entire
shape.
Compared to existing methods, this frees our network from being constrained to a
fixed genus, relying on a template, or requiring an extremely large collection
of shapes during training.
We demonstrate that even when trained on a single shape, our method can
generalize to novel meshes.
We design our network to encode vertex position data in a local frame ensuring
rotation and translation invariances without resorting to handcrafted predefined
feature descriptors.

We demonstrate the effectiveness of our method with a variety of qualitative and
quantitative experiments.
Our method generates subdivided meshes that are closer to the true
high-resolution shapes than traditional interpolatory and non-interpolatory
subdivision methods, even when trained with a small number of very dissimilar
exemplars.
We introduce a quantitative benchmark and show significant gains over classic
subdivision methods when measuring upsampling fidelity.
Finally, we show prototypical applications of Neural Subdivision to low-poly
mesh upsampling and 3D modeling.

\section{Related Work}\label{sec:relatedWork}
Our work builds directly upon the foundations of classic subdivision surfaces
and connects to the rapidly advancing field of neural geometry learning.
We focus this section on establishing context with past subdivision schemes and
contrasting our geometric learning contributions with contemporary works.

\subsection{Subdivision Surfaces}
The basic idea of subdivision is to ``define a smooth curve or surface as the
limit of sequence of successive refinements'' \cite{zorin2000subdivision}.
This broad definition admits a wide variety or ``zoo'' of different subdivision
schemes that would be outside the scope of this paper to cover thoroughly.
The history of subdivision surfaces reaches back to the early work on
irregular polygon meshes \cite{doo1978behaviour, Doo1978ASA} and the now
ubiquitous Catmull-Clark subdivision which produces quad meshes
\cite{catmull1978recursively}.
The linear method of \citet{loop1987smooth} for triangle meshes has reached
similar popularity, and is the basis for our non-linear neural subdivision.

Classic linear subdivision methods are defined by a combinatorial update
(splitting faces, adding vertices, and/or flipping edges \cite{Kobbelt00}) and a
vertex smoothing (repositioning step) based on local averaging of neighboring
vertex positions.
Subdivision methods are well studied from a theoretical perspective on the
existence, direct evaluation, and continuity of the limit surface
\cite{Stam98,zorin2007subdivision,karvciauskas2018new}.
Modelers typically manipulate a subdivision surface in a coarse to fine fashion.
Most modeling tools already visualize the limit surface or some
approximation of it, while the user manipulates the coarse level (cage) (see
\reffig{modeling}).
Beyond moving vertices, users can control the surface by adding creases (sharp
edges) \cite{hoppe1994piecewise,derose1998subdivision}.
Non-interpolating methods such as Catmull-Clark or Loop appear to be the most
popular, but interpolating methods do exist (e.g.,
\cite{dyn1990butterfly,kobbelt1996interpolatory,zorin1996interpolating}) and have similar
smoothness guarantees, although \emph{fairness} is harder to achieve (see
\reffig{teaser}).
Linear methods are easier to \update{analyze} and design to \update{guarantee} smoothness.
As a result, capturing details is left to the modeler or a
deterministic procedural routine (e.g.,
\cite{ToblerMW02,Tobler2002,velho2002algorithmic}).

\begin{figure}
    \centering
    \includegraphics[width=3.33in]{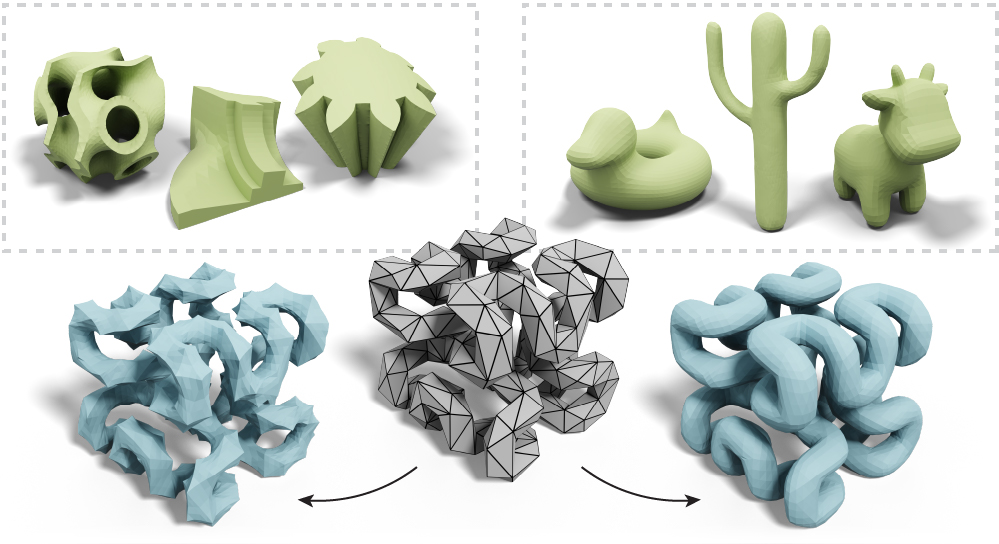}
    \caption{
        \update{Our subdivision is data-driven. Training on a set of mechanitical objects (left, green) or a set of smooth organic objects (right, green) leads to drastically different styles (blue). \textcopyright Gyroid Puzzl by eemmett (top left) and Hilbert Cube by tbuser (bottom) under CC BY-SA.}
        }
    \label{fig:mechE_vs_smooth}
    \vspace{-5pt}
\end{figure}
\update{Our neural subdivision acts similar to non-linear subdivision methods, with the subdivision rule in this case being a non-linear function learned by a neural network.}
Non-linear subdivision has been studied from the mathematical perspective
\cite{floater1997nonlinear,schaefer2008nonlinear} and also
as a mechanism to maintain certain geometric invariants during each level of
subdivision (e.g., circle-preserving \cite{sabin2004circle}, quad planarity
\cite{liu2006geometric,bobenko2019multi}, developability \cite{Tang2014Form,rabinovich2018shape},
Möbius-regularity \cite{vaxman2018canonical}, cloth wrinkliness
\cite{Ladislav2011cloth}).
One general approach is to combine a linear subdivision with an online
geometric optimization, \update{and recursively apply the non-linear rule an arbitrary, if not infinite number of times, akin to classic linear rules.}
Our approach can be viewed as an extreme form of precomputation, where the
optimization is the training procedure and the fixed network is applied
generally as a non-linear function evaluation.
\update{The choice of data in the training will influence the ``style'' of our non-linear subdivision (see \reffig{mechE_vs_smooth}).}
\update{Although our method is non-linear, it is trained to work well for a pre-specified finite number of times.}
% %

Recently, \citet{preiner2019gaussian} introduce\update{d} a new non-linear subdivision
method that treats the coarse shape probabilistically. Their contributions are
orthogonal to ours, \update{and while} we base our method on Loop subdivision, we could in theory
 extend our network to learn on top of this more powerful subdivision
method.

%!TEX root = ../LoD.tex
\begin{figure}
    \centering
    \includegraphics[width=3.33in]{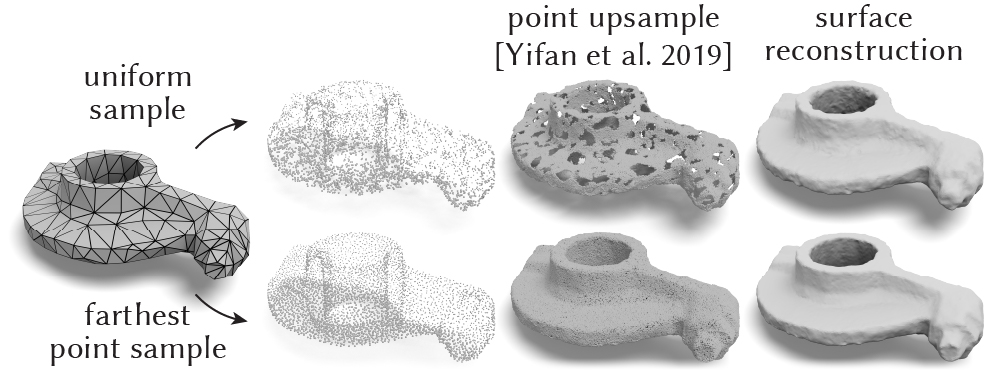}
    \caption{
        \update{One can use existing point upsampling methods to refine coarse meshes by (1) sampling, (2) upsampling \cite{Yifan_2019_CVPR}, and (3) reconstruction \cite{kazhdan2013screened}. However,} this may lead to artifacts since it lacks information about the surface, and requires the use of expensive surface reconstruction as a post-process.
        }
    \label{fig:ptUpsample}
    \vspace{-5pt}
\end{figure}
\begin{figure}
    \centering
    \includegraphics[width=3.33in]{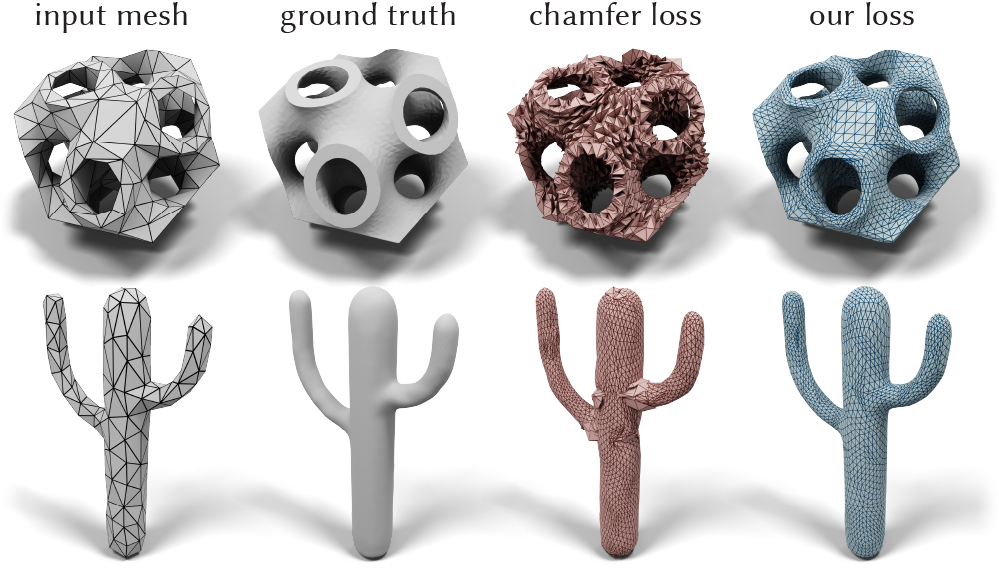}
    \vspace{-5pt}
    \caption{We compare the same model trained using (a) chamfer distance (which only measures error between point sets) and (b) our $\ell^2$ loss based on shape correspondences. The model trained using the chamfer distance fails to capture the surface topology (red). In contrast, our loss function leads to manifold output meshes (blue). \update{\textcopyright Gyroid Puzzle by emmett (top) under CC BY-SA.}}
    \label{fig:chamfer}
    \vspace{-10pt}
\end{figure}
\subsection{Neural Geometry Learning}

% applications
Recent advances in generative neural networks enabled the use of learnable components in 3D modeling applications
such as shape completion~\cite{Li_iccv19}, single-view~\cite{what3d_cvpr19} and multi-view~\cite{sitzmann2019deepvoxels} reconstruction, and modeling-by-parts~\cite{chaudhuri2020}.

\begin{figure*}
    \centering
    \includegraphics[width=7in]{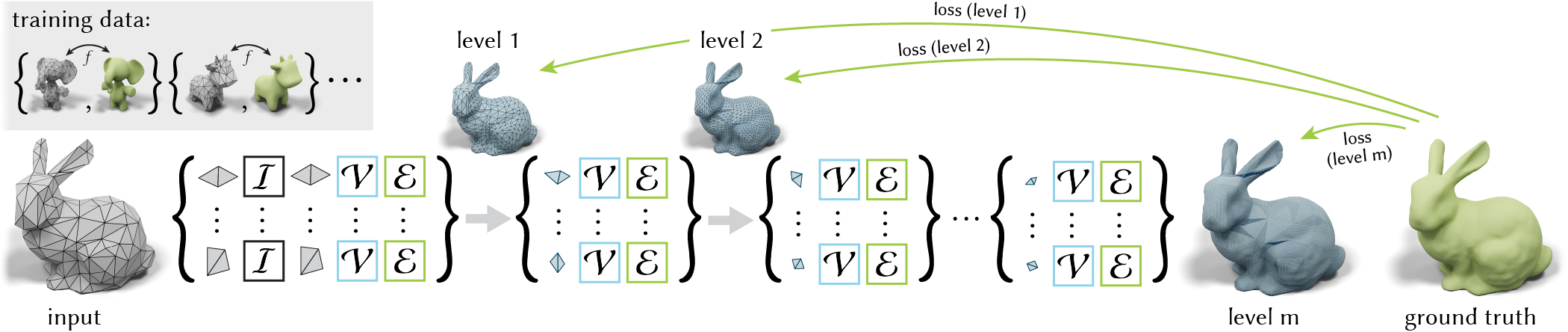}
    \caption{Neural subdivision takes a coarse triangle mesh (gray) as input and outputs a sequence of subdivided meshes (blue) with different levels of details. During training, we minimize the $\ell^2$ loss from the ground truth (green) to the output meshes (blue) across levels. Our training data consists of pairs of coarse and fine meshes (top left) with a bijective map $f$ between each pair. }
    \label{fig:overview}
    \vspace{-5pt}
\end{figure*}
The closest to our neural mesh subdivision application are the deep point cloud upsampling techniques~\cite{Yu18_PUNet,Li_iccv19,Yifan_2019_CVPR}.
The disadvantage of using a point cloud as input is that it lacks connectivity information, and requires the neural network to implicitly estimate the structure of the underlying manifold.
Meshes can also be more efficient at representing feature-less regions with larger planar elements, providing a wider reception field to our mesh-based neural network.
Mesh output is preferred for many standard graphics pipelines, thus, a post process is often required~\cite{kazhdan2013screened} to convert the output of point-based methods
to meshes, which prevents building an end-to-end trainable system. \reffig{ptUpsample} illustrates the output of a point upsampling method that was pre-trained on a collection of statues~\cite{Yifan_2019_CVPR} (see \refapp{pointUpsample} for implementation details).
%
%  \vova{I find it somewhat misplaced in related work. First, it's unclear how exactly we produced the result, so we'd need to add something like ``we sample XXX points on the coarse mesh, up-sample the points, apply poisson surface reconstruction to create the final mesh (?)'', which is kind of too wordy. Second, we also should show what coarse points and up-sampled points look like, otherwise it is not clear whether the artifacts are due to up-sampler's failures or something else (e.g., input is too sparse or PSR fails). Third, I don't think we should show our output next to Yifan's since the training data is different, so it's not really a proper comparison; instead, we can refer to teaser for people to look at our result. Of course, ideally, we want to have a proper comparison in results section.} \derek{done wth the figure update :) }
%

%
% mesh representations
Our work is related to other neural mesh generation techniques. Free-form generation of meshes as a set of vertices and faces is infeasible with current deep learning methods,
due to the lack of regular structure, uneven discretization, and combinatorial variability in the possible outputs, limiting such approaches to very coarse
outputs~\cite{dai2019scan2mesh}. A common alternative is to deform a global template either by predicting vertex coordinates~\cite{Tan18vaemesh,Ranjan18facemesh}
or by training a deformation network that warps the entire 3D domain conditioned on a latent vector that encodes the deformation target~\cite{Groueix18a,yifan2019neural}.
While these approaches usually produce meshes with higher resolution, their output is limited to deformations of a single shape.
Some techniques propose using generic templates such as spheres~\cite{Wang18pixel2mesh,Wen19pixel2mesh} or 2D atlases~\cite{Groueix18}, which place limitations
on the topology of the output. In contrast to these techniques, our method refines the mesh locally, and thus, respects the topology of the input (which could be arbitrary).
Another advantage of our local refinement approach is that we do not require co-aligned training data with a well-defined object space, the output of our subdivision networks is translation
and rotation invariant since it can be described in a local coordinate system of the input patch.

%
% mesh learning
There are several options for analyzing a mesh patch with a neural network, such as using a local~\cite{MasBosBroVan15,Poulenard2018geodesicNN} or global~\cite{Maron17}
parameterization to unfold a mesh into 2D grid, or apply graph-based techniques adapted for meshes~\cite{kostrikov2018surfacenetworks,Wang19b}
(see \cite{Bronstein16} for a more comprehensive survey).
Our approach is inspired by MeshCNN~\cite{hanocka2019meshcnn}. Their method directly learns filters over the local mesh structure via undirected edges, and shows applications in deterministic tasks. In contrast, we focus on generative tasks and develop a novel set of features over the \emph{half-flaps} -- an edge along with its two adjacent triangles. Each half-flap has a canonical orientation which gives us well-defined local frames which are crucial for our network's rigid motion invariance.

Geometry generation techniques are typically trained with reconstruction losses that measure how well does the generated surface approximate the known target.
Surface-to-surface distances are commonly employed, with correspondences defined via closest-point queries (aka chamfer distance)~\cite{barrow1977parametric, fan2017point}.
However, the closest-point approach matches many points to the same point, while leaving other points unmatched, resulting in self-overlaps and unrepresented areas (see \reffig{chamfer}).

Indeed, prior work demonstrates that using higher quality correspondence (e.g., ground truth mapping) significantly improves results~\cite{Groueix18a}. While the latter
is not available in our setting, we propose a data generation technique for creating various coarse variants of the same high-res mesh with a low-distortion bijective map.
Bijectivity is crucial for the quality of our training data, ensuring no self-overlaps exist and that every part of the target surface has a pre-image on the coarse mesh.

\vspace{-0.5pt}
\section{Neural Subdivision} \label{sec:overview}
In the following we overview \update{the} main components of our neural subdivision: the test-time inference pipeline, training and loss (\refsec{train}), data-generation (\refsec{train}), and finally \update{the} network architecture (\refsec{network}). \update{Later sections will discuss these components in detail.}

\paragraph{Inference.} As illustrated in \reffig{overview}, our method takes a coarse triangle mesh (gray) as input and recursively refines it by subdividing each triangle to create additional vertices and faces. The output is a sequence of subdivided meshes (blue) with different levels of details. Our subdivision process follows a simple topological update \update{rule (same as Loop)}, namely inserting new vertices at the midpoints of all edges. It then uses a neural network to predict new positions for all vertices, at each new level of subdivision.

\begin{figure*}
    \centering
    \includegraphics[width=7.0in]{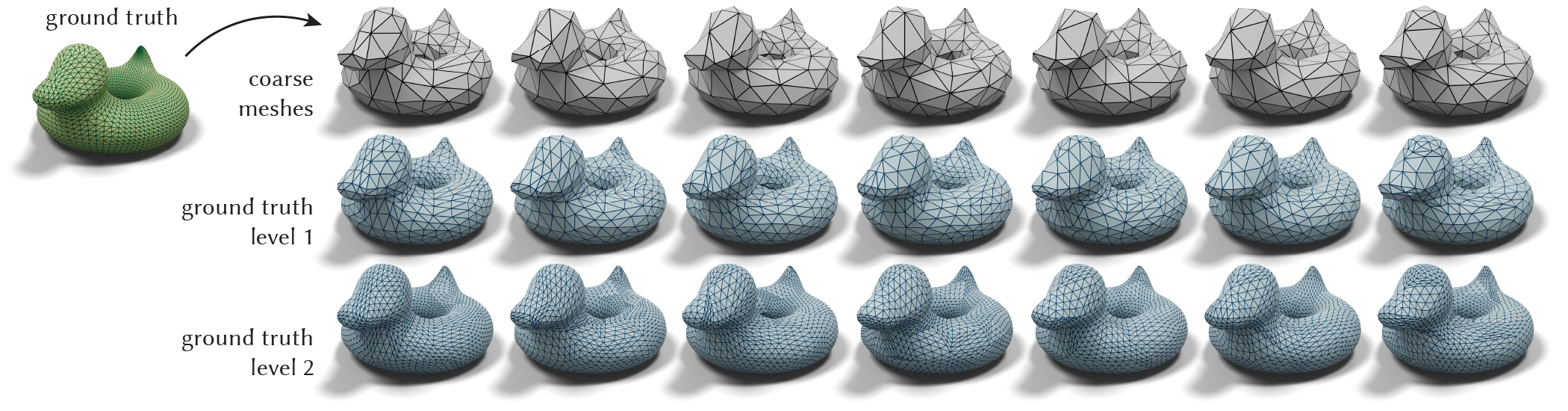}
    \caption{Given a ground truth shape (green), we use random edge collapses to create several coarse meshes (gray). For each coarse mesh, we subdivide the mesh and use the bijective map to determine the position on the ground truth for all the vertices across different levels. The blue meshes are the ground truth subdivisions that exhibit one-to-one vertex correspondences to the network predictions. }
    \label{fig:trainData}
    \vspace{-5pt}
\end{figure*}
\begin{figure}
    \centering
    \includegraphics[width=3.33in]{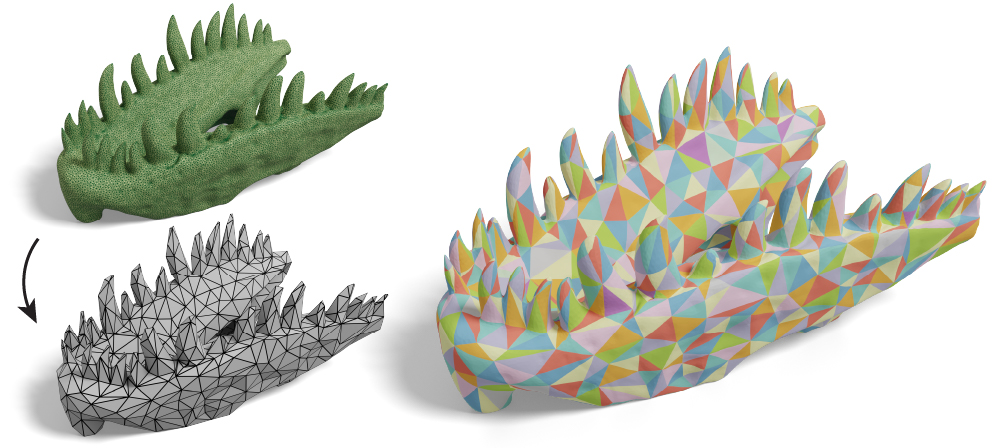}
    \caption{Given an edge collapse algorithm of choice, we plug in our successive self-parameterization described in \refsec{SSS} to compute a bijective map between the \update{original mesh (green) and its decimated version (gray).} We visualize the map by coloring the fine mesh using the triangulation of the coarse mesh (right). \update{\textcopyright Tarbosaurus Skull by gpvillamil under CC BY-SA.}}
    \label{fig:SSS}
\end{figure}
\paragraph{Training and loss.} The data we generate provides us with correspondences between predicted vertices and points \update{inside the triangle} on the ground truth shape\update{. We} train our network with the simple $ℓ^2$ loss, by measuring the distance between each predicted vertex position at every level of subdivision and its corresponding point on the \update{original} shape (green). As there is no existing dataset consisting of pairs of high-quality meshes and subdivision surfaces in correspondence, we instead develop a novel technique for generating training data, comprising of coarse and fine meshes with bijective mappings between them.

\paragraph{Data generation.} We first note that each vertex $v$ created from a subdivision step has a well-defined mapping back to the coarse mesh, defined by mapping that vertex to its corresponding midpoint. Thus, each \update{subdivided} mesh at any level of subdivision can be mapped back to the initial coarse mesh via a sequence of mid-point-to-vertex or vertex-to-vertex maps. In practice we use barycentric coordinates to encode this subdivided-to-coarse bijective mapping, $g$. Hence, if we had a bijective mapping $f$ between the coarse \update{mesh} and the \update{original mesh}, we could define a \update{unique point on the original mesh corresponding} to $v$, by compositing the two maps: $f\left(g\left(v\right)\right)$.

\update{Thus,} the only missing part is to create coarse and fine meshes with bijective mappings between them. We achieve this by taking a high-resolution training mesh and sequentially coarsening it by applying random sequences of edge collapses, thereby generating a sequence of coarsened meshes. We maintain low-distortion correspondences between the coarsened and original mesh by computing a conformal map between the 1-ring edge neighborhood (before the collapse) and the 1-ring vertex neighborhood (after the collapse). Composition of these maps \update{creates} a dense bijective map $f$ between the coarse and \update{original} meshes, \update{which is then directly applied to the training (\reffig{SSS}).}

\paragraph{Advantages of our training approach.} In comparison to closest point losses that are commonly used to train generative neural networks, our correspondence-based loss is aware of the manifold structure (\reffig{chamfer}) and is orders of magnitude faster to compute (\reffig{runtime}). Bijectivity and continuity of the map ensure that the entire ground truth surface is captured by some region of the coarse mesh (\reffig{point2mesh}). The low distortion encourages uniformity, which in turn enables the reproduction of the target surface with just a few uniform subdivisions, and, more importantly reduces the variance in the signal the network needs to learn. We can further leverage the low-distortion map to map an additional signal, such as texture (\reffig{texture}). As our training data contains many pairs with different random decimations of the same ground truth (\refapp{trainData}), our network is able to learn how to generalize across discretization.

\paragraph{Network architecture. } Similarly to the subdivision process, the learnable modules of our network are applied recursively. They operate over atomic local mesh neighborhoods and predict differential features (meaning they represent geometry in the local coordinates of the mesh, and not in world coordinates). These features are then used to compute vertex coordinates at the new level of subdivision. We define three types of modules applied at three sequential steps. During the \initialization step, we first compute differential per-vertex quantities that are based on the local coordinate frame. A learnable module \I is applied to the 1-ring neighborhood of every vertex to map these differential quantities to a high-dimensional feature vector stored at the vertex. Note that this high-dimensional feature vector is a concatenation of a learnable latent space which \emph{encodes} local geometry of the patch, and differential quantities which directly \emph{represent} local geometry and enable us to reconstruct the vertex coordinates.

\begin{wrapfigure}[6]{r}{1.33in}
	\raggedleft
    \vspace{-8pt}
	\hspace*{-0.7\columnsep}
	\includegraphics[width=1.45in, trim={6mm 0mm -1mm 0mm}]{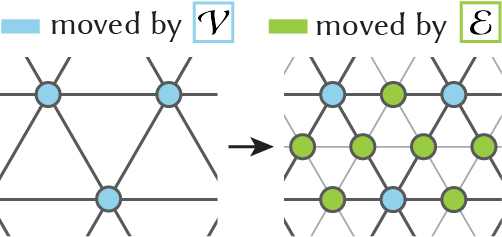}
\end{wrapfigure}
For each subsequent subdivision iteration, we assumes that the topology is updated following the Loop subdivision scheme, splitting each edge at midpoint, and consequently, subdividing each triangle into four (see inset). A \vertex step uses the module \V to predict vertex features for the next level of subdivision based on its 1-ring neighborhood, where vertices affected by this step only involve corners of triangles at the previous mesh level. Then, an \edge step uses the module \E to compute features of vertices added at midpoints based on the pair of vertices that were connected by an edge at the previous mesh level.

\begin{wrapfigure}[5]{r}{1.33in}
	\raggedleft
    \vspace{-12pt}
    % \vspace{-7pt}
	\hspace*{-0.7\columnsep}
	\includegraphics[width=1.45in, trim={6mm 0mm -1mm 0mm}]{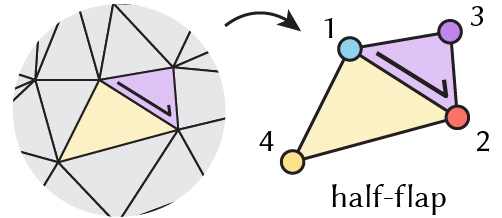}
	\label{fig:half_flap}
\end{wrapfigure}
Our modules share a very similar architecture and heavily rely on a learnable operator defined over \emph{a half-flap}: a directed edge and its two adjacent triangles (see the inset). We use the directed edge to define the local coordinate frame which is used to estimate the differential features of either input or output of learnable modules. %These are used both as inputs and outputs of all the steps.
Note also that the directed edge allows us to order the four adjacent vertices of the flap in a canonical way. We concatenate their features and feed them into shallow multi-layer perceptrons (MLP). The weights of the MLPs are shared within each module type and across all levels of subdivision. Both modules \I and \V process all half-flaps defined by an outgoing edge and use average pooling to combine the half-flap features into per-vertex features. The module \E also combines features from two half-flaps (both directions of the edge) via average pooling.
Since our architecture is local, and uses input and output features that are invariant to rigid motions, it exhibits an impressive ability to generalize from example, even when trained on a single fine mesh.

\section{Data Generation and Training} \label{sec:train}
While our network architecture and invariant layers are crucial for its ability to learn subdivisions, it by its own is only half of the two main components that together facilitate high-quality neural subdivisions.
The other half consists of the training process, data and the loss function.

\begin{figure}
    \centering
    \includegraphics[width=3.33in]{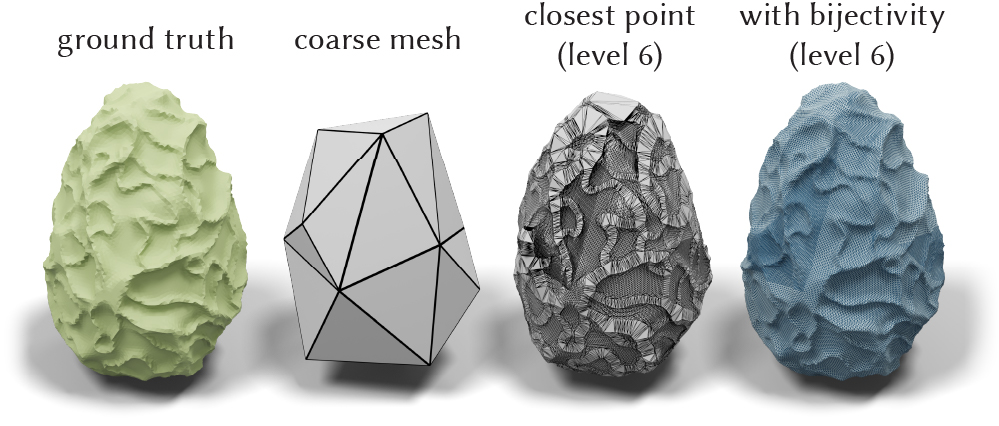}
    \caption{\update{Given a ground truth/coarse mesh pair, naively using ``closest-point-on-mesh'' to estimate correspondences between the level-6 subdivided mesh and the ground truth} results in a non-bijective map, causing the loss function to fail to capture the entire ground truth mesh (third column). Our successive self-parameterization ensures bijectivity, which implies the entire ground truth surface will be captured (right)}
    \label{fig:point2mesh}
    \vspace{-5pt}
\end{figure}

Consider a naive approach to the subdivision training: generate pairs of coarse/fine meshes by a decimation algorithm; measure the distance between the network's predicted subdivision and the ground truth, for instance by the average distance between predicted points and their projections on the ground truth mesh; iterate over coarse/fine pairs while optimizing the loss.
This naive approach has a major caveat. Computing correspondences using the chamfer-like loss (\reffig{chamfer}) or point-to-mesh distance (\reffig{point2mesh}) is known to lead to incorrect and self-overlapping matches between shapes.
This leads to a badly training set up because the loss itself exhibits artifacts.
\begin{figure}
    \centering
    \includegraphics[width=3.33in]{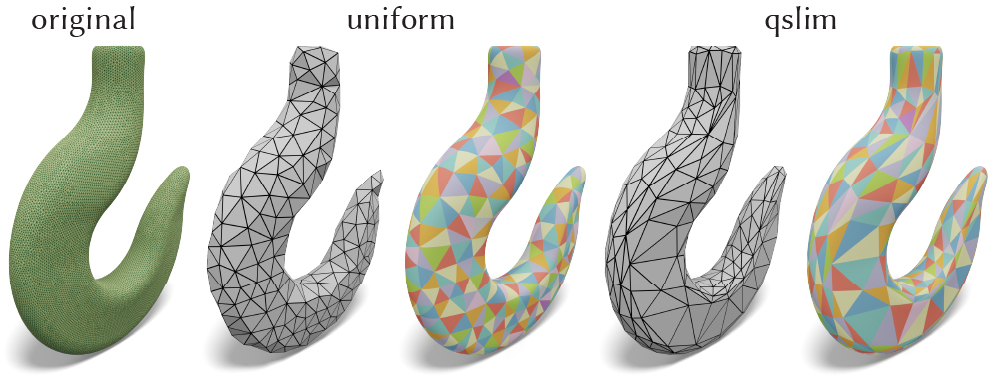}
    \caption{Different edge-collapse algorithms can be used in a plug-and-play manner to create, for instance, a uniform-area parameterization (middle) and an appearance-preserving parameterization (right). This flexibility is used to create training data with diverse types of discretizations.}
    \label{fig:controlEC}
    \vspace{-5pt}
\end{figure}

In lieu of this naive approach, we consider the fact that a pair of coarse and fine meshes both approximate the same underlying smooth surface.
This motivates us to compute the correspondences based on the intrinsic geometry, instead of an ad-hoc correspondence.
The outcome is a high-quality bijective map between each pair of coarse and fine meshes, enabling us to obtain one-to-one vertex correspondences. Therefore a simple $\ell^2$ loss is sufficient to correctly measure the error between every level of neural subdivision and the ground truth shape.

% In order to train our neural network we require data, namely many pairs of coarse and fine meshes, with a high-quality bijective map for each pair. For any given point on the simplified mesh (including an arbitrary point within a face), the bijective map enables us to retrieve its corresponding point on the input mesh, and vice versa. This is crucial for the formulation of our loss, which relies on one-to-one vertex correspondences. We therefore devise an algorithm named successive self-parameterization for creating many coarse and fine mesh pairs with high-quality bijective maps, designed specifically for creating data to train neural subdivision.

\subsection{Successive Self-Parameterization} \label{sec:SSS}
One possible solution is to apply general shape matching techniques. But ensuring bijectivity in general shape matching is difficult. For instance, it requires the two shapes to have the same number of vertices \cite{vestner2017product}, or a user-guided common domain \cite{Schreiner04, Praun2001Consistent}, or user-provided landmark correspondences \cite{kraevoy2004cross, aigerman2014lifted, aigerman2015seamless} (see \cite{van2011survey} for a survey). However, our problem is considerably simpler, since we aim to construct a map between different discretizations of the same shape, and we have full control on the decimation procedure.

\begin{figure}
    \centering
    \includegraphics[width=3.33in]{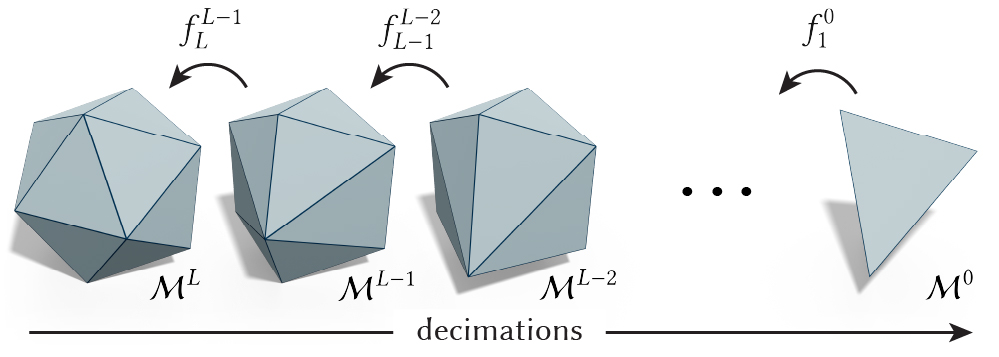}
    \caption{We compute a bijective map for each edge collapse. The bijective map from the coarsest mesh $\M^0$ to the input mesh $\M^L$ is then computed by composing all the maps $f^{l-1}_l$.}
    \label{fig:composition}
    \vspace{-5pt}
\end{figure}
The closest solution to our problem is a seminal work -- \MAPS \cite{lee1998maps} -- on self-parameterization. Given an input mesh, \MAPS computes the bijective map by successively removing vertices of the maximum independent set. \update{Since then, several improvements have been proposed~\cite{Guskov2000,Guskov2002, Khodakovsky2003}}. Unfortunately, \update{they} cannot be directly applied to edge collapses for creating training data for our learning task (see \refapp{compareMAPS}). We need an algorithm that has the flexibility to be used with any edge decimation method, so that we can generate a diverse collection of coarse meshes (see \reffig{controlEC}). Fortunately, the idea from \cite{Cohen1997, cohen1998appearance, CohenMO03} for minimizing mesh/texture deviation leads us to generalize the idea of \MAPS to any edge collapses.

Our method for computing the bijective map, designed specifically for creating data to train neural subdivision, combines the idea of self-parameterization from \MAPS \cite{lee1998maps} and the idea of successive mapping from \cite{Cohen1997, cohen1998appearance,CohenMO03}. Thus, we call it \emph{successive self-parameterization}.
This combination enables us to compute the parameterization intrinsically to avoid the requirement of having a given UV map, such as in the method of \citet{LiuFerguson2017SSE}. The result of the combination is extremely simple. It is a two-step module that can be applied to any choice of edge-collapse algorithm (see \reffig{controlEC}) and it will output a bijective map after the decimation. Hence, the inputs to successive self-parameterization are a triangle mesh and an edge collapse algorithm of choice, and the output is a decimated mesh with a corresponding bijective map between the input and the decimated model. For the sake of reproducibility, we reiterate the core ideas from \cite{lee1998maps, Cohen1997, cohen1998appearance}, and describe how to combine both ideas.

We denote the input triangle mesh as $\M^L = (\mV^L, \mF^L)$, where $\mV^L, \mF^L$ are vertex positions and face information respectively at the original level $L$. The input mesh $\M^L$ is successively simplified into a series of meshes $\M^l = (\mV^l, \mF^l)$ with $0 ≤ l ≤ L$, where $\M^0 = (\mV^0, \mF^0)$ is the coarsest mesh. For each edge collapse $\M^l → \M^{l-1}$, we compute the bijective map $f^{l-1}_l: \M^{l-1} → \M^l$ (see \reffig{composition}) on the fly. The final map $f^0_L: \M^0 → \M^L$ is computed via composition,
\begin{align} \label{equ:composition}
    f^0_L =   f^{L-1}_L ∘ \cdots ∘ f^0_1 .
\end{align}
We now focus our discussion on the computation of a bijective map for a single edge collapse.

\begin{figure}
    \centering
    \includegraphics[width=3.33in]{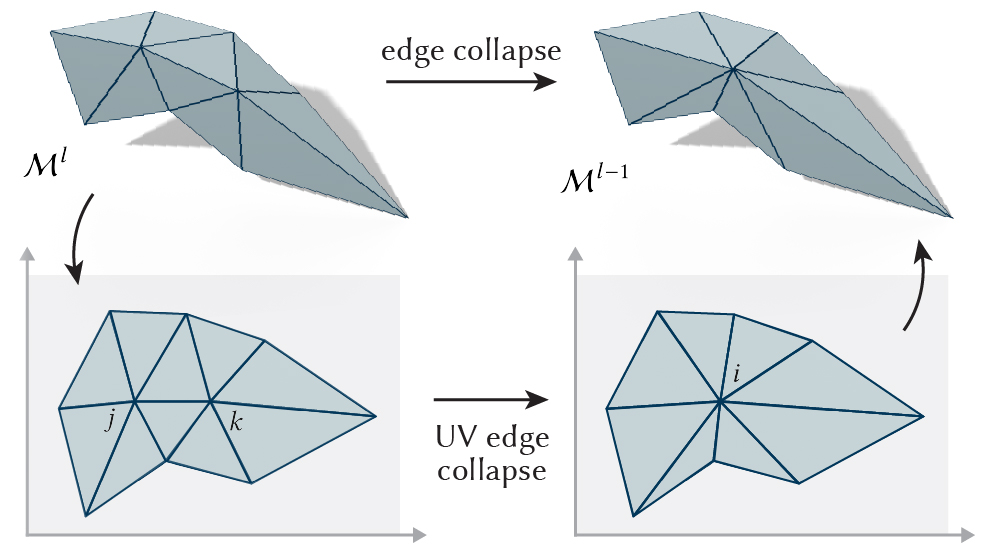}
    \caption{For each edge collapse, we simultaneously collapse the edge on the 3D mesh (top) and the UV domain (bottom). As the boundary vertices of the edge's 1-ring are preserved through the edge collapse, we constrain the flattened boundary in UV space to be at the same position when computing the conformal parameterization of the post-collapse 1-ring.}
    \label{fig:edgeCollapse}
    \vspace{-5pt}
\end{figure}
\begin{figure}
    \centering
    \includegraphics[width=3.33in]{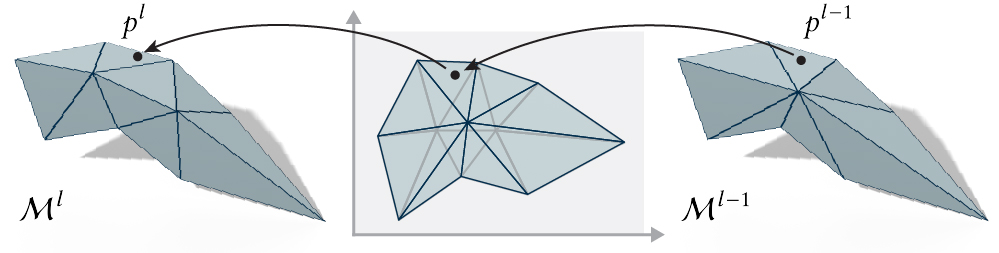}
    \caption{Since both the pre-collapse and post-collapse parameterizations of the 1-ring map it into the same 2D domain, we can easily use the shared UV space to map a point back and forth between $\M^l$ and $\M^{l-1}$.}
    \label{fig:baryQuery}
    \vspace{-5pt}
\end{figure}
\subsection{Single Edge Collapse}
In each edge collapse, the triangulation remains the same, except for the neighborhood of the collapsed edge. Let $\N(i)$ be the neighboring vertices of a vertex $i$ and let $\N(j,k) = \N(j) \cup \N(k)$  denote the neighboring vertices of an edge $(j,k)$. After each collapse, the algorithm computes the bijective map for the edge's 1-ring $\N(j,k)$, in two stages. It first parameterizes the neighborhood $\N(j,k)$ (prior to the collapse) into 2D. It then performs the edge collapse both on the 3D mesh, and in UV space, as depicted in \reffig{edgeCollapse}. The key observation from \cite{Cohen1997, cohen1998appearance} is that the boundary vertices of $\N(j,k)$ before the collapse become the boundary vertices of $\N(i)$ after the collapse. Hence the UV parameterization of the 1-ring remains valid and injective after the collapse. Then, for any given point $p^{l-1} ∈ \M^{l-1}$ (represented in barycentric coordinates), we can utilize the shared UV parameterization to map $p^{l-1}$ to its corresponding barycentric point $p^{l} ∈ \M^{l}$ and vice-versa, as shown in \reffig{baryQuery}.

\begin{figure}
    \centering
    \includegraphics[width=3.33in]{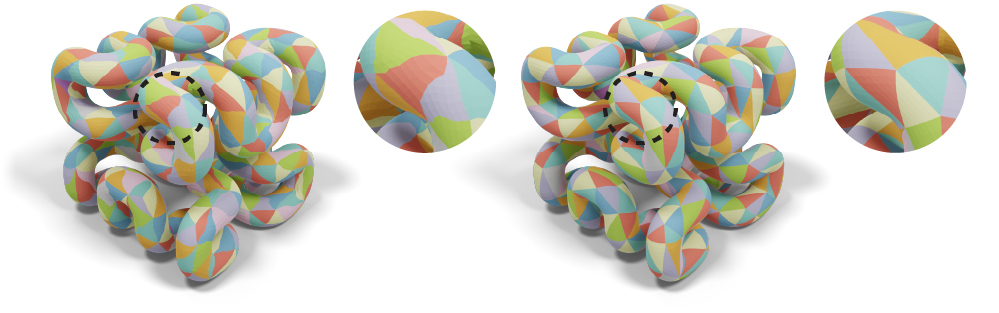}
    \caption{Using a different parameterization technique that does not result in a conformal flattening leads to a distorted parameterization (left), in contrast to the conformal parameterization we use, that reduces the amount of angle distortion accumulated throughout the edge collapse sequence (right). \update{ \textcopyright Hilbert Cube by tbuser under CC BY-SA.}}
    \label{fig:withWithoutConformal}
    \vspace{-5pt}
\end{figure}
Following the idea of \MAPS \cite{lee1998maps}, we use conformal flattening \cite{mullen2008spectral} to compute the UV parameterization of the 1-rings, \reffig{edgeCollapse}. After collapsing an edge and inserting the new vertex $\vv ∈ \R^3$ , we determine this vertex's UV location by performing another conformal flattening with fixed boundary. The conformality of the map is crucial, as it minimizes angle distortion which would otherwise accumulate throughout the successive parameterizations, leading to distorted, skewed correspondences and hindered learning of the network (see \reffig{withWithoutConformal}).

\subsection{Implementation}
\begin{figure*}
    \centering
    \includegraphics[width=7.0in]{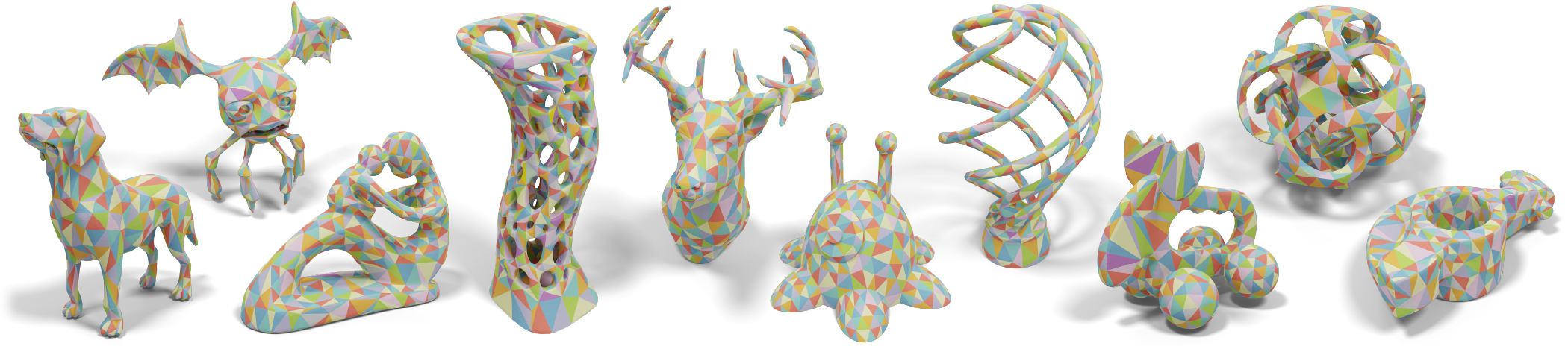}
    \caption{Checking the criteria of collapsible edges is crucial for the robustness of the successive self-parameterization. \update{From left to right, \textcopyright Psycho by Aeva (2nd, CC BY-SA), Parametric Sculpture by MCompeau (4th, CC BY-NC), Deer Head by TakeshiMurata (5th, CC BY-SA), Brain Slug by Zarquon (6th, CC BY-NC-SA), Spiral Light Bulb by benglish (7th, CC BY-SA), and Metratron by addy (9th, GNU).}}
    \label{fig:robustSSS}
    \vspace{-5pt}
\end{figure*}
\update{Successive self-parameterization can be used with any edge collapse algorithm simply by adding two additional steps} (see \refapp{SSSImplement}). The actual edge collapse algorithm, such as \qslim \cite{garland1997surface}, takes $\mathcal{O}(N\log N)$ time, and the flattening is a constant cost on top of each collapse (assuming valence is bounded). Thus the complexity of the entire algorithm containing both edge collapses and successive self-parameterization is still $\mathcal{O}(N\log N)$.

The robustness of the parameterization algorithm \update{relies heavily} on the robustness of the underlying edge collapse algorithm. Edge collapses that may lead to self-intersections can result in unusable maps. In \refapp{criteria}, we summarize our criteria for checking the validity of an edge collapse. This is crucial to ensure that we can generate training data using a wide range of shapes (see \reffig{robustSSS}).

\subsection{Training Data \& Loss Computation}
\begin{wrapfigure}[13]{r}{0.4\linewidth}
    \vspace{-12pt}
	\hspace*{-0.5\columnsep}
    \begin{minipage}[b]{1.45in}
    \includegraphics[width=1.45in, trim={6mm 0mm 0mm 0mm}]{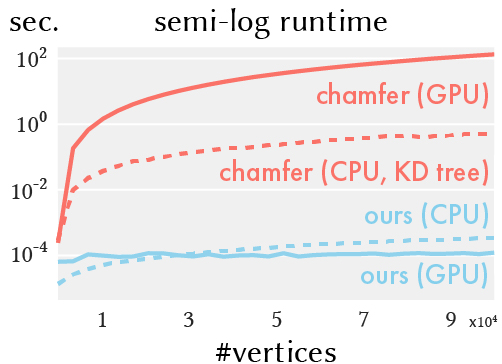}
    \vspace*{-1.5\intextsep}
    \caption{Our loss computation is orders of magnitude faster than the chamfer loss on the GPU (\textsc{Kaolin}~\cite{kaolin2019arxiv}) or the CPU (our KD-tree-based implementation).}
    \label{fig:runtime}
    \end{minipage}
\end{wrapfigure}
Our training data is constructed by applying the successive self-parameterization on top of random edge collapses. In \reffig{trainData}, given a high-resolution shape (green), we use \qslim \cite{garland1997surface} with a random sequence of edge collapses to construct several different decimated models (gray). During the collapse, we plug in our self-parameterization to obtain a high-quality bijective map for each coarse and fine pair.

After the network subdivides the coarse mesh, we use the map to retrieve one-to-one correspondences to the input shape. Specifically, when retrieving the correspondences, we use the Loop topology update to add points in the middle of each edge, \update{e.g., the point with barycentric coordinates \mbox{$(0.5, 0.5, 0)$} in a triangle of the coarse mesh. We use these barycentric coordinates $\vb$ on the coarse mesh to obtain the barycentric coordinates $f(\vb)$ on the fine mesh, as illustrated in \reffig{baryQuery} using the bijective map $f$. During training, suppose $\mathcal{E}(\vb)$ is the vertex position output by the network $\mathcal{E}$. We measure the per-vertex loss with the $\ell^2$ distance \mbox{$\| f(\vb) - \mathcal{E}(\vb) \|_2$}.}
%
% Once we have the one-to-one correspondences, we simply use the point-wise $\ell^2$ norm as the loss function for training.
%
Compared to the chamfer distance~\cite{barrow1977parametric}, a widely used distance in training 3D generative models \cite{fan2017point}, our loss computation is orders of magnitude faster (see \reffig{runtime}).
\section{Network Architecture}\label{sec:network}
\begin{figure*}
    \centering
    \includegraphics[width=7.0in]{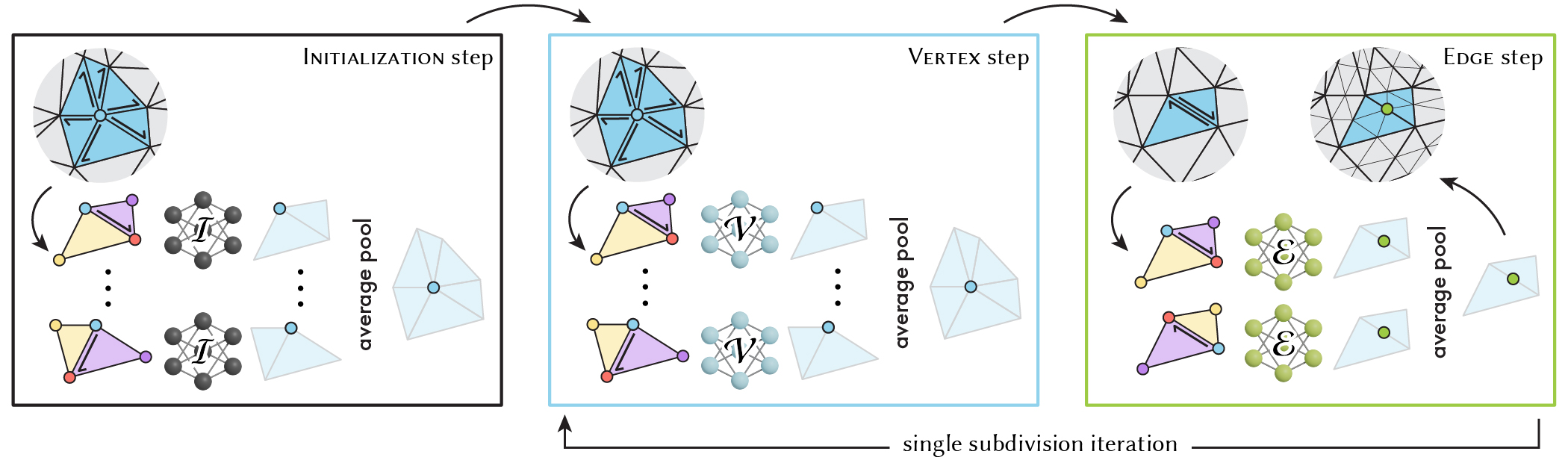}
    \caption{Our neural subdivision consists of three sequential steps: \initialization, \vertex, and \edge, with three network modules: $\mathcal{I}$, $\mathcal{V}$, and $\mathcal{E}$ for each step respectively. In both \initialization and \vertex steps, we apply \V and \E for the half-flaps of all the outgoing edges of a vertex, and use average pooling to combine the output features back to the center vertex (blue). In the \edge step, we apply \E to both half-flaps of an undirected edge and use average pooling to map the output features to the center vertex (green) of the edge.}
    \label{fig:pipeline}
    \vspace{-5pt}
\end{figure*}
Given a mesh at a previous level of subdivision along with a known topological update \update{rule} (mid-point subdivision as used by Loop), our neural network computes all vertex coordinates for the subdivided mesh. Our process involves three main steps illustrated in \reffig{pipeline}. \update{The} \initialization step uses \update{a} learnable neural module \I to map input per-vertex features to high-dimensional feature vector at each vertex. In each subdivision iteration, \update{the} \vertex step uses \update{a} learnable module \V to update features at corners of triangles of the input mesh, and \update{the} \edge step uses \update{a} learnable module \E to compute features of vertices that were generated at mid-points of edges of the input mesh. \update{Our network is inspired by classical subdivision algorithms which have two sets of rules: to update (1)~\emph{even} vertices from previous iterations, and (2)~the newly inserted \emph{odd} vertices. One difference of our approach is that we apply \V and \E in sequence, instead of in parallel. This allows us to harness neighborhood information from previous steps.}

We make several design choices that are critical to the ability of our network to generalize well even from very small amount of training data.
First, even though all mesh update steps are global (i.e., they affect every vertex of the mesh), our learnable modules that are used in these steps operate over local mesh patches and share weights. Thus, even a single training pair provides many local mesh patches to train our neural modules.
Second, our modules operate over original discrete elements of the mesh, and do not require re-parameterizing or re-sampling the surface. Representing input and output using the mesh discretization enables us to preserve the topology of the input, and generalize to novel meshes with different topology.
Third, we represent our vertices using differential quantities with respect to a local coordinate frame instead of using global coordinates. Thus our neural modules operate over representation that is invariant to rigid motion which simplifies training and improves their ability to generalize.

The key component of our neural module is a learnable operator that takes \emph{half-flap}, a 2-face flap adjacent to a half-edge, inspired by the edge convolution approach of \citet{hanocka2019meshcnn}. We choose to use half-flap (instead of a flap around an undirected edge) since it provides a unique canonical orientation for the four vertices at the corners of adjacent faces. It also provides a well-defined local coordinate frame which we will use to define differential vertex quantities for the input and output (see the inset). Each flap operator is a shallow multi-layer perceptron (MLP) defined over features of four ordered points. We train one operator per module ($\mathcal{I}$, $\mathcal{V}$, $\mathcal{E}$) across all levels of subdivision and training examples.

Equipped with the half-flap operator, we use average pooling to aggregate features from different half-flaps to per-vertex features in all our neural subdivision steps. \initialization and \vertex steps apply the half-flap operator to every outgoing edge in a 1-ring neighborhood of a vertex, and average pooling aggregates per-half-flap outputs into a per-vertex feature. \edge step only considers per-vertex features at two endpoints of a subdivided edge to compute the feature of the inserted vertex. Thus, it simply applies half-flap operator for each directions of the edge and again uses average pooling to get the vertex feature.
\begin{figure}
    \centering
    \includegraphics[width=3.33in]{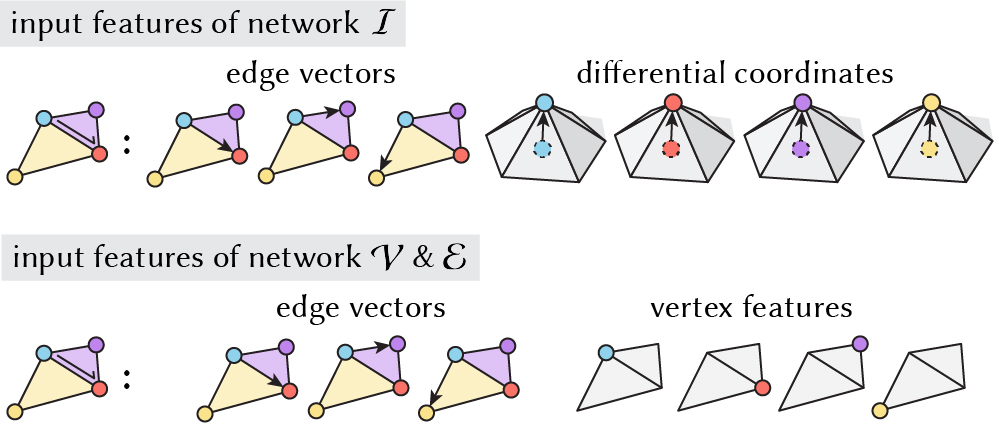}
    \caption{The input feature to module \I consists of three edge vectors from the source vertex (blue) and vectors of the differential coordinates for the four vertices. The input features to module \V and \E are three edge vectors with per-vertex high-dimensional features from the previous steps.}
    \label{fig:flapInput}
    \vspace{-5pt}
\end{figure}
\begin{figure}
    \centering
    \includegraphics[width=3.33in]{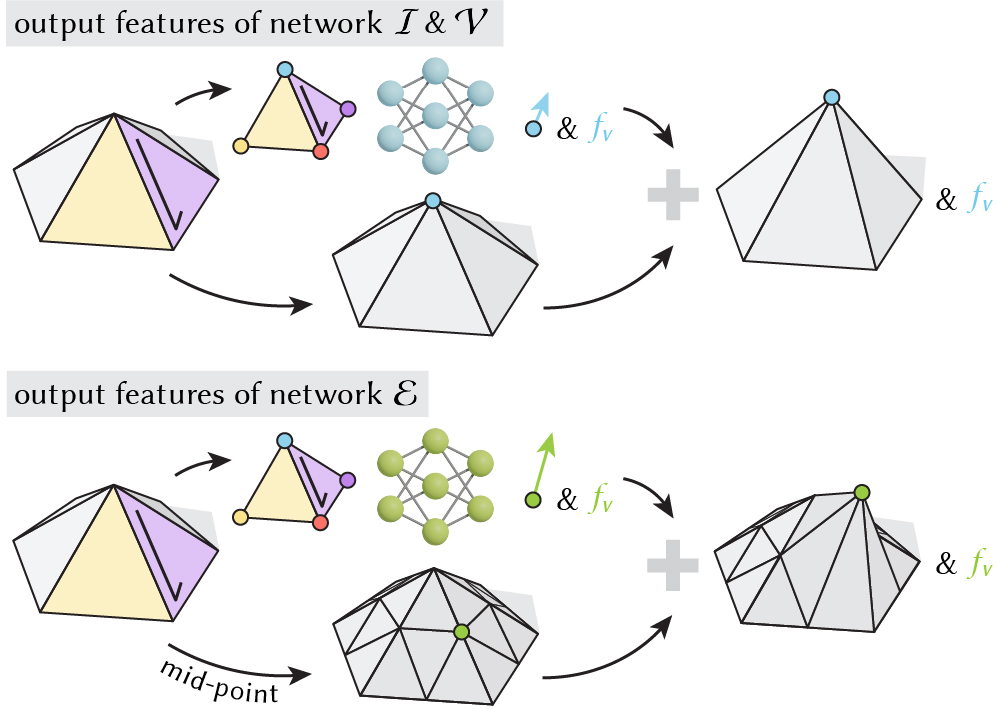}
    \caption{The outputs of modules \I and \V are the displacement vector from the starting vertex and a learned feature vector $f_v$ stored at the source vertex (blue). The outputs of the module \E are the displacement from the edge mid-point (green) and the feature $f_v$ stored at the mid-point.}
    \label{fig:flapOutput}
    \vspace{-5pt}
\end{figure}

\begin{wrapfigure}[5]{r}{1.33in}
	\raggedleft
    \vspace{-12pt}
	\hspace*{-0.7\columnsep}
	\includegraphics[width=1.45in, trim={6mm 0mm -1mm 0mm}]{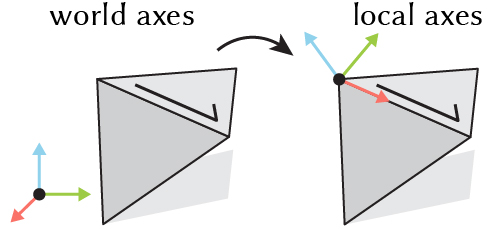}
	\label{fig:localFrames}
\end{wrapfigure}
The final critical element of our architecture design is the representation for the input and output. As mentioned before, we use local differential quantities to ensure invariance to rigid transformation. The input features for the half-flaps used in \initialization step by module \I consist of three edge vectors (originating at the source vertex of half-flap) and differential coordinates of each vertex, as illustrated in \reffig{flapInput}, top. The vector of differential coordinates stores the discrete curvature information and is defined as the difference between the absolute coordinates of a vertex and the average of its immediate neighbors in the mesh \cite{sorkine2005laplacian}. To achieve rotation invariance we represent our differential quantities in the local frame of each half-flap (see the inset), where we treat the half-edge direction as the x-axis, the edge normal computed via averaging the two adjacent face normals as the z-axis, and the cross product of the previous two axes becomes our y-axis.
The input to half-flap operators used in \vertex and \edge steps is similar (\reffig{flapInput}, bottom), where we use edge vectors and per-vertex high-dimensional learned features (either produced by \initialization step or by previous subdivision iteration).
The output of half-flaps used in \vertex and \edge steps includes high-dimensional learned features and differential quantities that can be used to reconstruct the vertex position.
For the latter we use the vertex displacement vector from the mid-point subdivided mesh (see \reffig{flapOutput}) in the local coordinate system of the half-flap. For the \initialization and the \vertex networks, the predicted displacements live on the vertices; for the \edge network, the predicted displacements live on the edge midpoints. In our experiments, we notice there is no difference between predicting from the mid-point subdivided surface or other subdivision surfaces (see \refapp{ablation}), so we choose mid-point subdivision for simplicity.
We estimate global coordinates of vertices after each step to visualize intermediate levels of subdivision and compute the loss function, and convert global coordinates to local differential per-vertex quantities before each step to ensure that each network only observes translation- and rotation- invariant representations.
\begin{figure}
    \centering
    \includegraphics[width=3.33in]{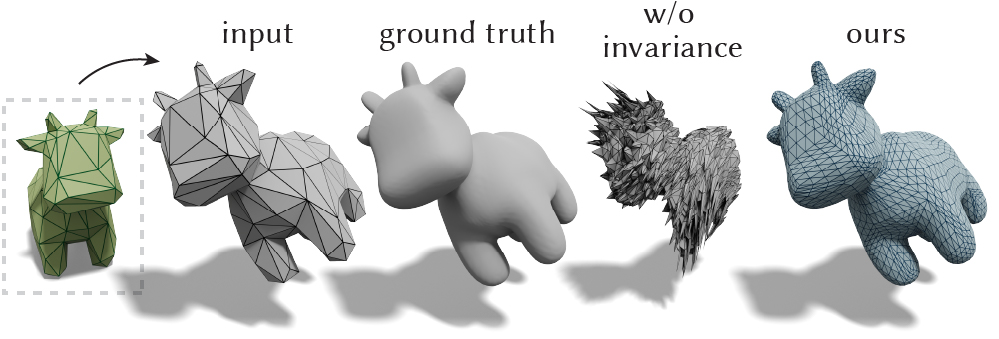}
    \caption{We use differential quantities stored in the local frames as our inputs and outputs. This design makes our network invariant to rigid motions and significantly boosts the quality compared to an approach without invariance.}
    \label{fig:toy_translation}
    \vspace{-5pt}
\end{figure}

\begin{table}[t]
    \setlength{\tabcolsep}{5.425pt}
    \centering
    \caption{Hyperparameters of our sub-networks. All networks are fully-connected multi-layer perceptrons with two hidden layers.}
    \vspace{-5pt}
    \begin{tabularx}{0.75\linewidth}{l|ccc}
        \toprule
        \textit{ } & \textit{network \I} & \textit{network \V} & \textit{network \E} \\
        % \midrule
        \rowcolor{derekTableBlue}
        $\textit{f}_\text{in}$  & $3⋅3 + 4⋅3$ & $3⋅3 + 4⋅32$ & $3⋅3 + 4⋅32$ \\
        $\textit{fc}_1$   & 32 & 32 & 32  \\
        \rowcolor{derekTableBlue}
        $\textit{fc}_2$   & 32 & 32 & 32 \\
        $\textit{f}_\text{out}$  & $3+29$ & $3+29$ & $3+29$  \\
        \bottomrule
    \end{tabularx}
    \smallskip
    \label{tab:network_params}
\end{table}
\reffig{toy_translation} illustrates that invariant representation is critical to the quality of results. We demonstrate that even when trained on an identical true shape, a slight rigid motion of that shape renders learned weights completely inapplicable at inference time. We also observe that incorporating the differential coordinates as part of the input features makes the training converge faster (see \refapp{ablation}). Thanks to our local half-flap operators and invariant representations we can train our architecture even with shallow 2-layer MLPs (see \reftab{network_params} for network hyper-parameters). We further evaluate other design decisions and conclude that details such as whether to predict displacements from the mid-point or the Loop subdivision, whether to recursively apply the module \V, whether to measure loss across all levels, and whether to use input features proposed in \cite{hanocka2019meshcnn} offer small improvements to the convergence (see \refapp{ablation} for details).

\update{We implemented our network in \textsc{PyTorch} \cite{paszke2017automatic}. We use \textsc{ReLu} activation \cite{nair2010rectified}, and the \textsc{ADAM} optimizer \cite{kingma2014adam} with learning rate $0.002$.}
\section{Evaluations}\label{sec:results}
% So far we have discussed all the required ingredients, including the loss function, the training data, and the network architecture.
We evaluate our neural subdivision with a range of results of increasing complexity. We start by showing that we can generalize to isometric deformations, non-isometric deformations, shapes from different classes, and shapes from different types of discretizations. We summarize the details of our experiments in \refapp{experiments}.

\begin{figure}
    \centering
    \includegraphics[width=3.33in]{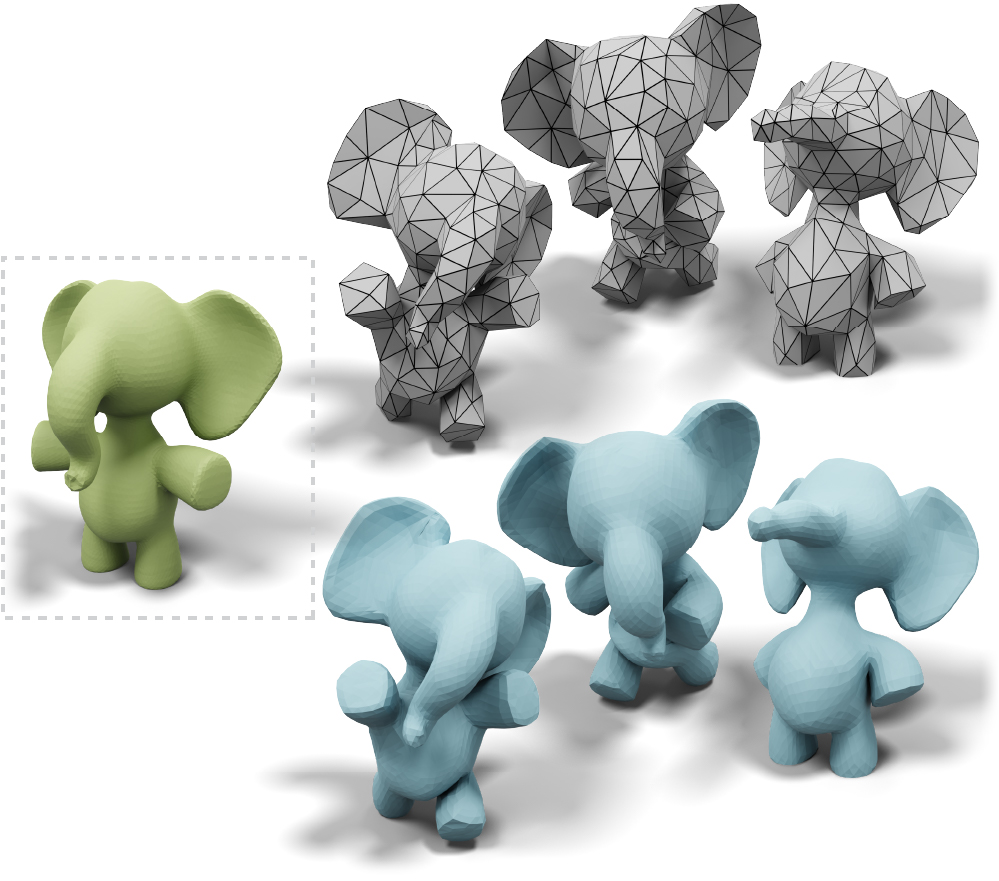}
    \caption{We train our network on a single pose (green) and the network is able to generalize to unseen poses (blue).}
    \label{fig:poses}
    \vspace{-5pt}
\end{figure}

In practice, modelers often manipulate the coarse subdivision cage of a character into different poses, and then apply the subdivision operator. This scenario implies that being able to train on one single pose and generalize to unseen poses is important for character animation. In \reffig{poses}, we train on a single pose (in green) and show that our network can generalize to unseen poses under (approximately) isometric deformations.

\begin{figure}
    \centering
    \includegraphics[width=3.33in]{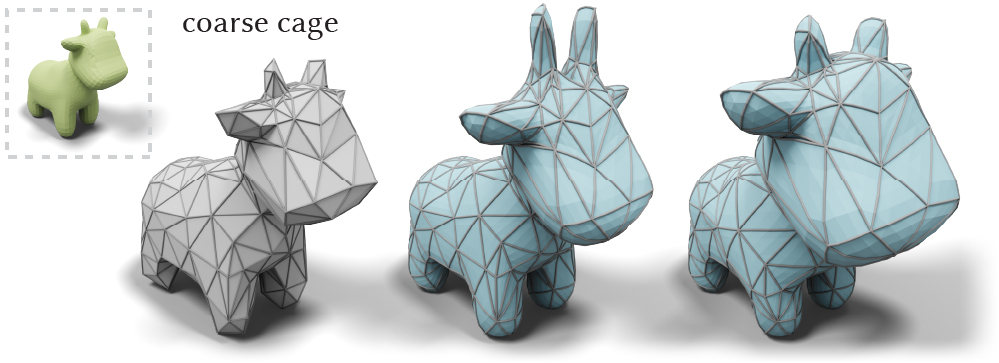}
    \vspace{-5pt}
    \caption{We mimic the modeling scenario by applying non-isometric deformations to the coarse cage (gray). Our subdivision network is able to generalize to unseen non-isometric deformations.}
    \label{fig:modeling}
    \vspace{-10pt}
\end{figure}
In addition to poses, in \reffig{modeling} we mimic the real scenario to manually change the coarse cage and show that the learned subdivision can also generalize to non-isometric deformations.

\begin{figure}
    \centering
    \includegraphics[width=3.33in]{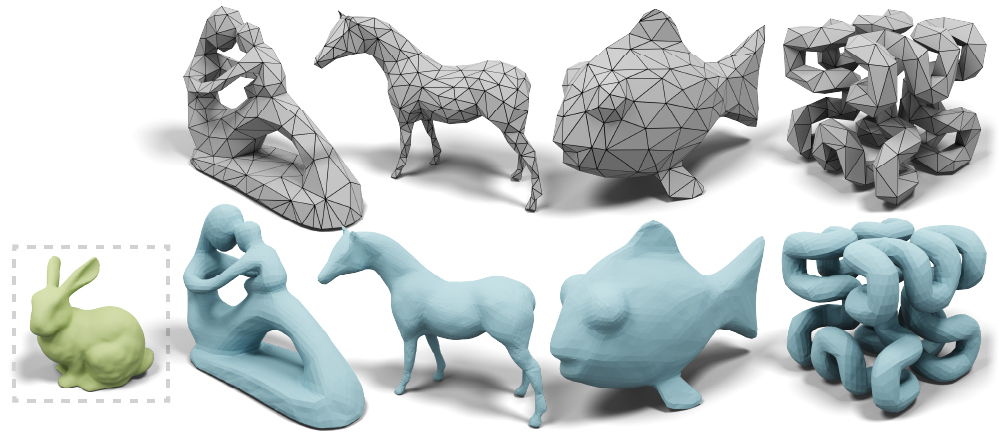}
    \caption{Even when trained on only a single shape (green bunny), our network can generalize to subdividing different geometries (blue). \update{ \textcopyright Hilbert Cube by tbuser (right) under CC BY-NC.}}
    \label{fig:singleShape}
    \vspace{-5pt}
\end{figure}

Subdivision operators are often used to create novel 3D content, which implies the importance of generalizing to totally different shapes. In \reffig{singleShape} we show that even when trained on only a single shape (green), our network is able to generalize to many other shapes (blue). \update{We also show that our network trained on classic Loop subdivision sequences is able to reproduce Loop subdivision on unseen shapes (\refapp{Loop}).}

\begin{figure}
    \centering
    \includegraphics[width=3.33in]{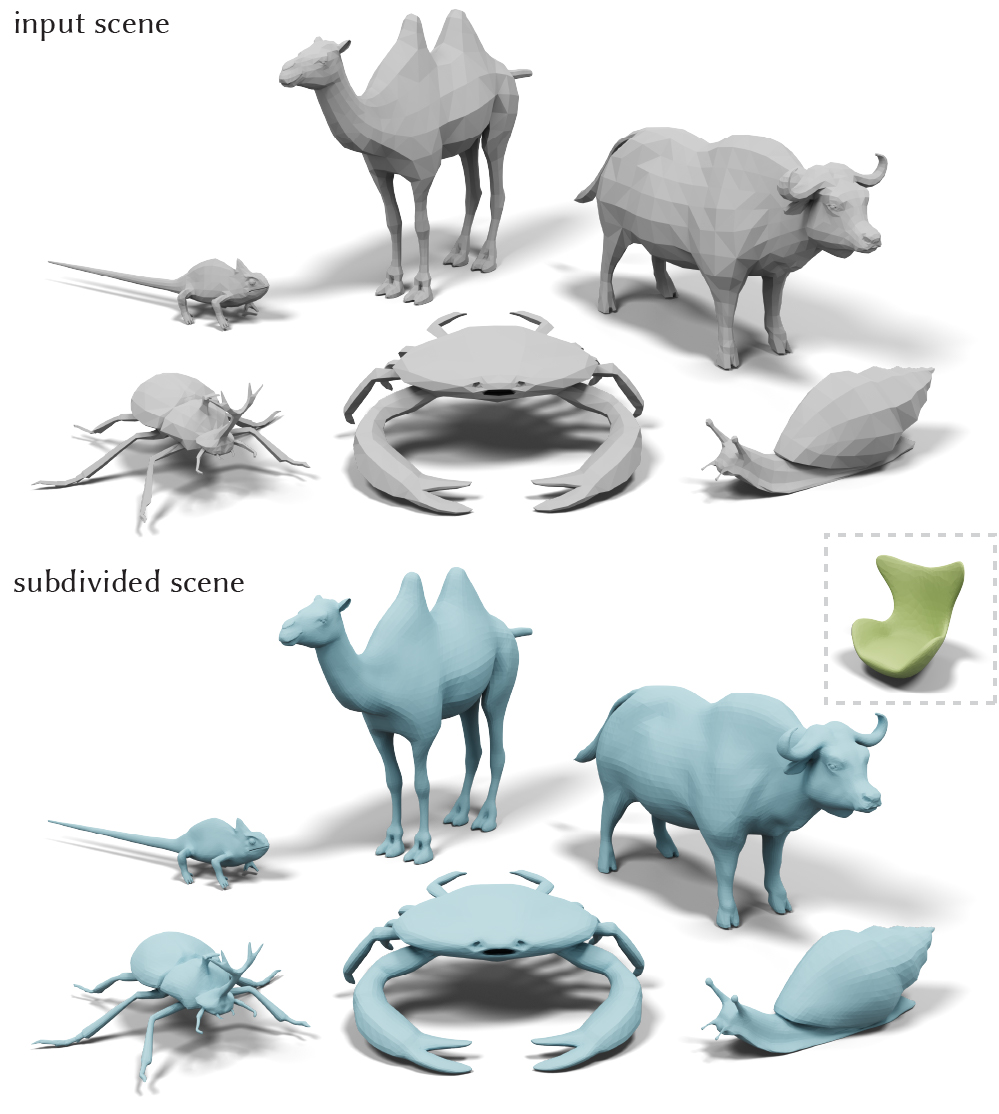}
    \caption{In addition to subdividing meshes constructed via decimation, our network can also generalize to subdivide meshes created by artists.}
    \label{fig:subdivScene}
    \vspace{-5pt}
\end{figure}
%
%Not just different shapes,
\update{We further evaluate neural subdivision on shape discretizations created in a totally independent way.} In \reffig{subdivScene} we obtain coarse shapes created by artists, instead of from edge collapses, and show that neural subdivision can still generalize well.

\begin{figure}
    \centering
    \includegraphics[width=3.33in]{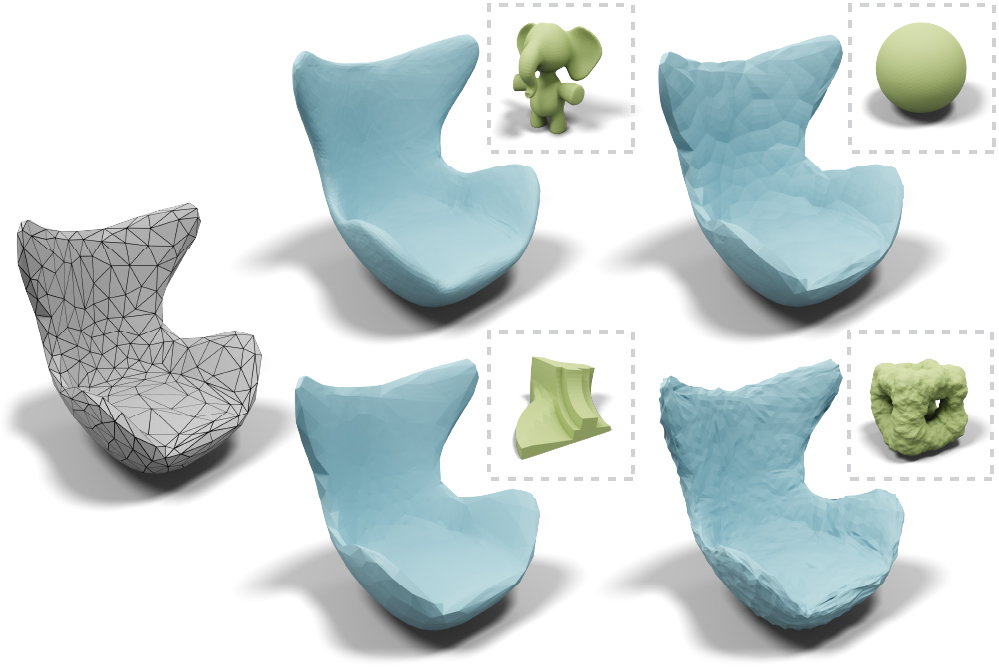}
    \caption{Using different shapes in training leads to stylized subdivision results (blue) biased towards the training shapes (green). \update{\textcopyright Egg Chair by TeamTeamUSA (left) under CC BY.}}
    \label{fig:dataDriven}
    \vspace{-5pt}
\end{figure}

\begin{figure}
    \centering
    \includegraphics[width=3.33in]{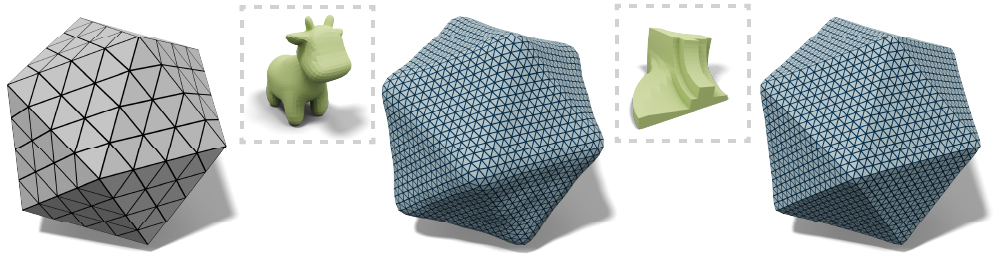}
    \caption{Training on a smooth shape leads to a smoother subdivision result (middle). Training on a man-made object can preserve the sharp creases (right). }
    \label{fig:lim_sharp}
    \vspace{-5pt}
\end{figure}

The ability to generalize even when trained on a single shape gives us the opportunity to do stylized subdivisions. In \reffig{dataDriven} our neural subdivision operators are aware of the ``style'' of the training shape and are able to create different results from the same coarse geometry. In \reffig{lim_sharp}, we show different results when trained on a smooth organic shape vs a man-made object with sharp contours.

To quantitatively analyze how our network generalizes to unseen shapes, we take the TOSCA dataset \cite{bronstein2008numerical} which contains 80 shapes with 9 categories to perform quantitative analysis. For the top table of \reftab{quantitative}, we train on a single category (\emph{Centaur}) and test on the remaining categories. \update{Our test shapes are generated by coarsening source meshes with \qslim down to 350-450 vertices. We measure the error between the two-level subdivided mesh and the original shape using Hausdorff distance, as well as mean surface distance} computed by the \textsc{metro} \cite{cignoni1998metro}. Our method consistently produces smaller errors compared to the classic Loop \cite{loop1987smooth} and modified butterfly \cite{zorin1996interpolating} subdivisions.

\begin{figure}
    \centering
    \includegraphics[width=3.33in]{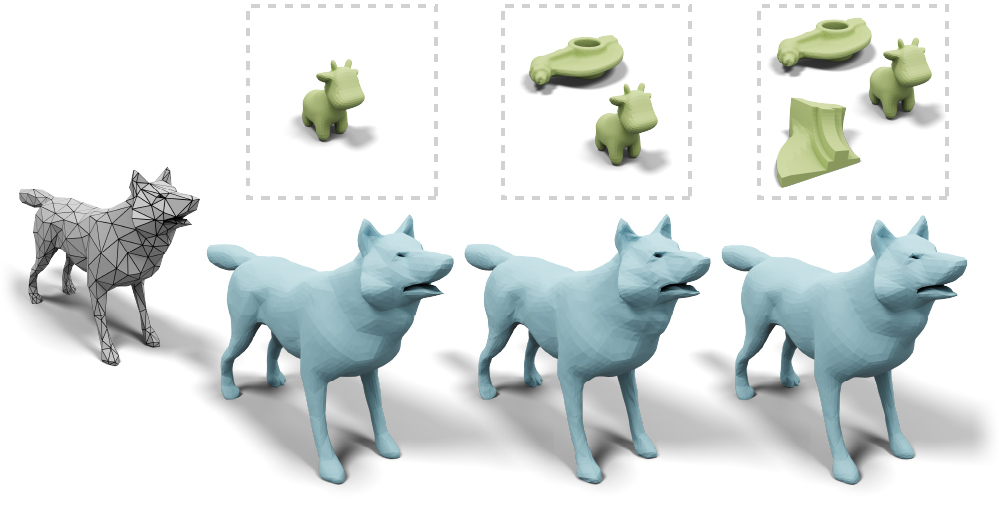}
    \caption{We \update{train our subdivision network on a mixture of organic and non-organic shapes}. We observe that training on more objects does not significantly change visual quality in this case.}
    \label{fig:moreShapes}
    \vspace{-5pt}
\end{figure}

We further evaluate our method when trained on multiple shapes and categories.
In \reffig{moreShapes}, we train the network on a increasing number of objects and observe that the results are visually similar. But our quantitative analysis in the bottom table of \reftab{quantitative} shows that training on more categories (\emph{Centaur, David, Horse}) can slightly reduce the error.
\begin{table}
    \setlength{\tabcolsep}{5.425pt}
    \centering
    \caption{
        We train on a single category, \emph{Centaur} (top table), and three categories, \emph{Centaur, David, Horse} (bottom table), separately, and evaluate by subdividing the rest of the TOSCA shapes. The results indicate that neural subdivision outperforms classic Loop subdivision \cite{loop1987smooth} and modified butterfly subdivision \cite{zorin1996interpolating} on two popular metrics: Hausdorff distance $\mathbb{H}$, and mean surface distance $\mathbb{M}$ computed via \textsc{metro} \cite{cignoni1998metro}.
    }
    % \vspace{-5pt}
    \begin{tabularx}{0.96\linewidth}{l|ccc|ccc}
        \toprule
        \textit{Category} & $\mathbb{H}_\text{loop}$ & $\mathbb{H}_\text{m.b.}$ & $\mathbb{H}_\text{ours}$ & $\mathbb{M}_\text{loop}$ & $\mathbb{M}_\text{m.b.}$ & $\mathbb{M}_\text{ours}$ \\
        \rowcolor{derekTableBlue}
        Cat   & 2.75 & 2.17 & \textbf{2.08} & 0.73 & 0.21 & \textbf{0.17}\\
        David & 2.95 & 2.13 & \textbf{1.83} & 0.88 & 0.27 & \textbf{0.20}\\
        \rowcolor{derekTableBlue}
        Dog  & 3.26 & 2.32 & \textbf{2.11} & 0.84 & 0.31 & \textbf{0.26}\\
        Gorilla & 4.53 & 3.17 & \textbf{2.56} & 1.27 & 0.48 & \textbf{0.36}\\
        \rowcolor{derekTableBlue}
        Horse & 5.87 & 4.53 & \textbf{4.04} & 1.51 & 0.50 & \textbf{0.45}\\
        Michael  & 3.88 & 2.71 & \textbf{2.24} & 1.12 & 0.38 & \textbf{0.28}\\
        \rowcolor{derekTableBlue}
        Victoria & 4.25 & 3.01 & \textbf{2.36} & 1.12 & 0.39 & \textbf{0.30}\\
        Wolf & 2.83 & 1.74 & \textbf{1.63} & 0.69 & 0.23 & \textbf{0.21}\\
        \bottomrule
    \end{tabularx}
    \quad\\ \quad\\
    \smallskip
    \ \begin{tabularx}{0.96\linewidth}{l|ccc|ccc}
        \toprule
        \textit{Category} & $\mathbb{H}_\text{loop}$ & $\mathbb{H}_\text{m.b.}$ & $\mathbb{H}_\text{ours}$ & $\mathbb{M}_\text{loop}$ & $\mathbb{M}_\text{m.b.}$ & $\mathbb{M}_\text{ours}$ \\
        \rowcolor{derekTableBlue}
        Cat   & 2.75 & 2.17 & \textbf{2.09} & 0.73 & 0.21 & \textbf{0.16}\\
        Dog  & 3.26 & 2.32 & \textbf{2.12} & 0.84 & 0.31 & \textbf{0.25}\\
        \rowcolor{derekTableBlue}
        Gorilla & 4.53 & 3.17 & \textbf{2.89} & 1.27 & 0.48 & \textbf{0.34}\\
        Michael  & 3.88 & 2.71 & \textbf{2.15} & 1.12 & 0.38 & \textbf{0.27}\\
        \rowcolor{derekTableBlue}
        Victoria & 4.25 & 3.01 & \textbf{2.49} & 1.12 & 0.39 & \textbf{0.28}\\
        Wolf & 2.83 & 1.74 & \textbf{1.65} & 0.69 & 0.23 & \textbf{0.20}\\
        \bottomrule
    \end{tabularx}
    \label{tab:quantitative}
\end{table}

\section{Limitations \& Future Work}
\begin{figure}
    \centering
    \includegraphics[width=3.33in]{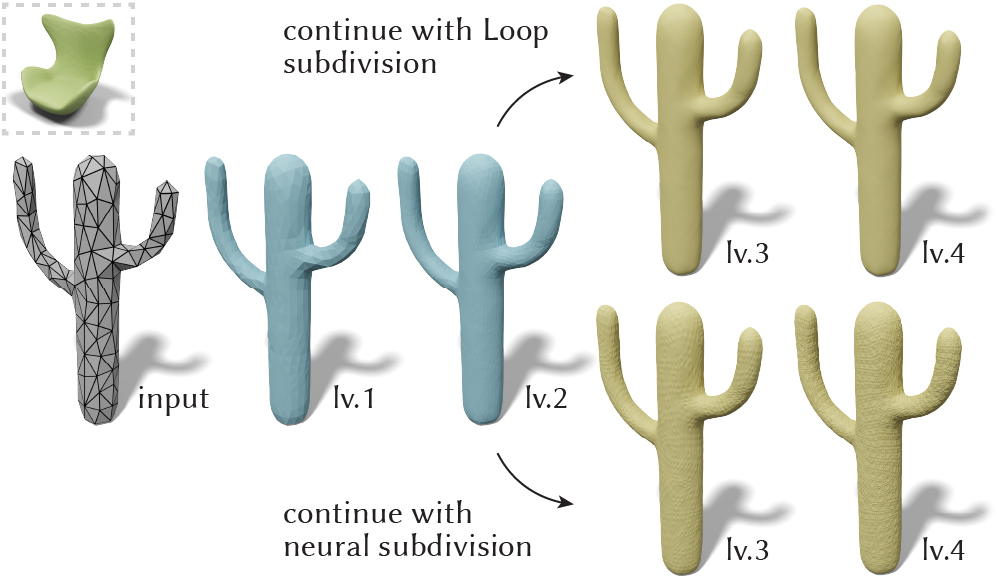}
    \caption{\update{Since our method induces a non-linear subdivision, there is no guarantee for the existence of a limit surface (bottom).}
    %Our method does not converge to a limit surface (bottom).
    An alternative is to apply neural subdivision at the trained levels, and continue with classic subdivision (top) \update{to ensure a smooth limit surface.}}
    \label{fig:lim_beyondSubd}
    \vspace{-5pt}
\end{figure}

\begin{figure}
    \centering
    \includegraphics[width=3.33in]{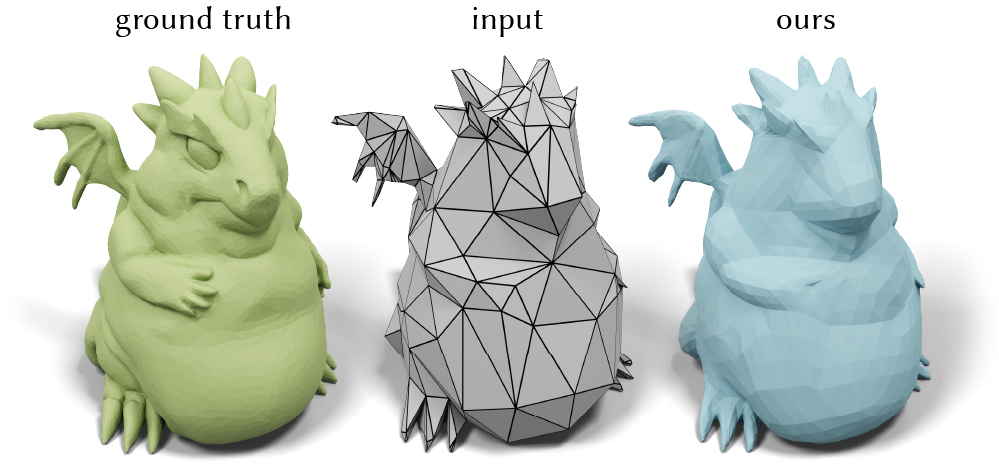}
    \caption{Our approach is based \update{on local geometry, and thus fails} to hallucinate semantic features. \update{\textcopyright Bratty Dragon by Splotchy Ink under CC BY.}}
    \label{fig:lim_semantic}
    \vspace{-5pt}
\end{figure}

Extending the neural subdivision framework to quadrilateral meshes \update{and surface with boundaries} would be closer to real-world modeling scenarios.
Making neural subdivision \update{scale-invariant} and converge to a limit surface (see \reffig{lim_beyondSubd}) are also desirable in practice.
Incorporating global information in the training could help the network \update{hallucinate} semantic features (see \reffig{lim_semantic}).
Applying architectures (e.g., Recurrent Neural Net) that are more suitable for sequence predictions could help the network to harness information from a wider neighborhood and to dive to a deeper subdivision level.
Training on data that contain a wide range of triangle aspect \update{ratios} and curvature information could further improve the robustness of the network.
Since our data-generation algorithm is extremely efficient, it \update{could be naturally used} in an online-learning setting, where our algorithm constantly draws new randomly-coarsened meshes on-the-fly. This can be extremely useful in, e.g., a GAN setting.
\update{As a first step towards neural subdivision, we showed reconstruction of fine meshes from coarse ones. Fully-fledged super-resolution, detail hallucination, and surface stylization are interesting next steps.}
All of these questions provide interesting topics for the future research on neural subdivision.

\begin{acks}
\update{Our research is funded in part by New Frontiers of Research Fund (NFRFE–201), the Ontario Early Research Award program, NSERC Discovery (RGPIN2017–05235, RGPAS–2017–507938), the Canada Research Chairs Program, the Fields Centre for Quantitative Analysis and Modelling and gifts by Adobe Systems, Autodesk and MESH Inc. We thank members of Dynamic Graphics Project at the University of Toronto; Thomas Davies, Oded Stein, Michael Tao, and Jackson Wang for early discussions; Rahul Arora, Seungbae Bang, Jiannan Li, Abhishek Madan, and Silvia Sell\'{a}n for experiments and generating test data; Honglin Chen, Eitan Grinspun, and Sarah Kushner for proofreading. We thank Mirela Ben-Chen for the enlightening advice on the experiments and the writing; Yifan Wang for running comparisons; and Ahmad Nasikun for evaluations. We obtained our test models from \textsc{Thingi10K} \cite{Thingi10K} and we thank all the artists for sharing a rich variety of 3D models. We especially thank John Hancock for the IT support which helped us smoothly conduct this research.}
\end{acks}

\bibliographystyle{ACM-Reference-Format}
\bibliography{reference.bib}
\appendix

\section{Implementation of Point Cloud Upsampling} \label{app:pointUpsample}
An alternative way to upsample a mesh is to first convert the mesh into point cloud via sampling over the surface, run point cloud upsampling algorithms, and then perform a surface reconstruction to convert the upsampled point cloud back to a mesh.
However, this procedure is expensive to incorporate into the interactive graphics pipeline, fails to produce surfaces with different levels of detail (see \reffig{allLevels}), and it fails to preserve textures (see \reffig{texture}). In addition, many non-trivial design decisions such as the number of samples to use and how to sample the surface would influence the quality of the results.
For example in \reffig{ptUpsample}, we first sample 5000 points with uniform and farthest point sampling, followed by the method of \citet{Yifan_2019_CVPR} pre-trained on statues to upsample the point cloud by 16$\times$, and then use the screened poisson reconstruction \cite{kazhdan2013screened} to reconstruct the surface.
In the figure we show that different sampling methods lead to different results. The lack of connectivity information also results in some surface artifacts.

\section{Implementation of Successive Self-Parameterization} \label{app:SSSImplement}
Incorporating successive self-parameterization only requires adding two additional local conformal parameterizations to the edge collapse algorithm of choice. Suppose we want to collapse an edge $(j,k)$. We first flatten the edge's 1-ring $\N(j,k)$,
% before collapsing the edge,
then we collapse the edge, then we perform another conformal flattening on the 1-ring $\N(i)$ of the newly inserted vertex $i$ after the collapse, with the boundary held to place from the previous flattening. This yields a bijective map with small computational cost because each flattening only involves a 1-ring (assuming the vertex valence is bounded).
% Our \textsc{MATLAB} implementation decimates the Stanford Bunny from 7000 vertices down to 500 vertices in 40 seconds, from which 19 seconds are spent on computing the map, with the remaining 21 seconds spent on edge collapses.

%
\begin{figure*}
    \centering
    \includegraphics[width=7.0in]{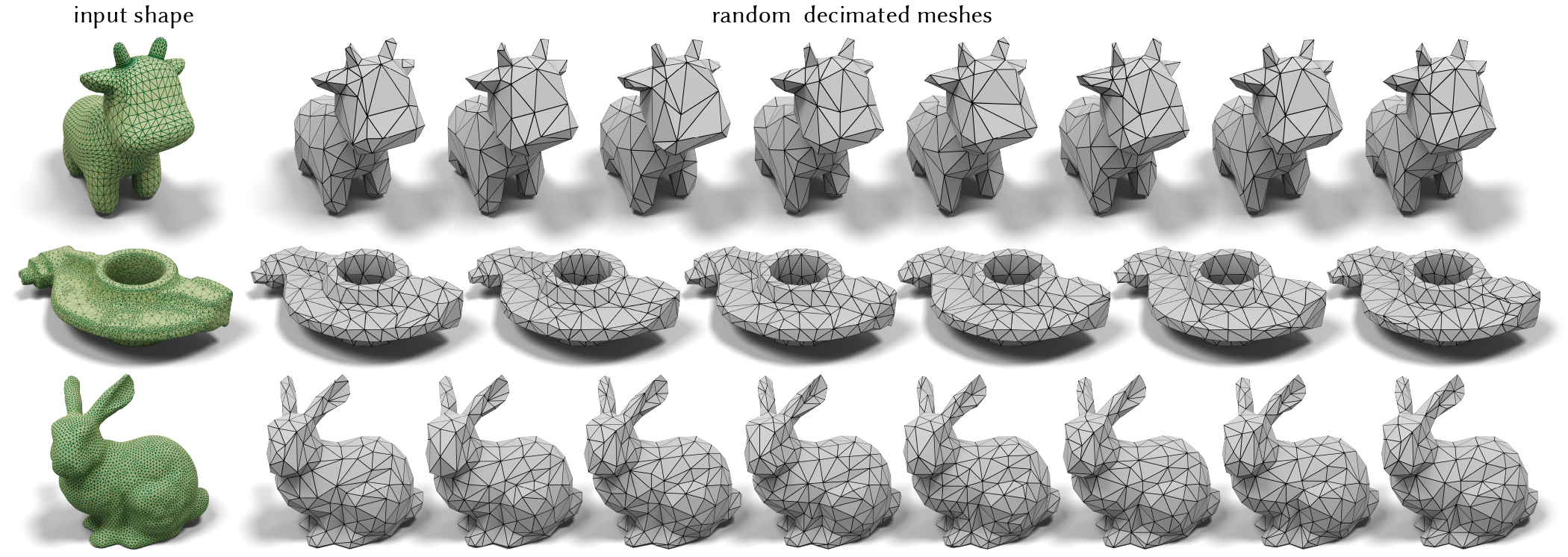}
    \caption{We perform \qslim with a random sequence of edge collapses to create different coarse discretizations (gray) from a single ground truth mesh (green).}
    \label{fig:manyShapes}
    \vspace{-5pt}
\end{figure*}
\section{Criteria for Collapsible Edges} \label{app:criteria}
During edge collapses, many issues such as flipped faces and non-manifold edges may appear. Resolving these issues is crucial to the robustness of successive self-parameterization (see \reffig{robustSSS}). We summarize our criteria for checking the validity of an edge collapse. If invalid, we simply avoid collapsing the edge at that iteration.

\paragraph{Euclidean face flips}
\begin{wrapfigure}[5]{r}{1.33in}
	\raggedleft
    \vspace{-10pt}
	\hspace*{-0.7\columnsep}
	\includegraphics[width=1.45in, trim={6mm 0mm -1mm 0mm}]{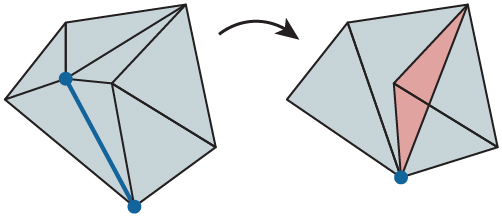}
	\label{fig:flipFace}
\end{wrapfigure}
Certain faces in the Euclidean space may suffer from normal flips after an edge collapse. To prevent flipped faces, we simply compare the unit face normal $\vn$ of each neighboring face $f_i$ before and after the collapse
\begin{align}
   {\vn}_{f_i}^\text{before} ⋅ {\vn}_{f_i}^\text{after} > δ.
\end{align}
Our default $δ = 0.2$ which is sufficient to avoid face flips in all our experiments.

\paragraph{UV face flips}
Flipped faces may also appear in the UV space due to both the conformal flattening and the edge collapse. We simply check whether the signed area of each UV face is positive before and after collapses to prevent having UV face flips.

\paragraph{Overlapped UV faces}
\begin{wrapfigure}[5]{r}{1.33in}
	\raggedleft
    \vspace{-10pt}
	\hspace*{-0.7\columnsep}
	\includegraphics[width=1.45in, trim={6mm 0mm -1mm 0mm}]{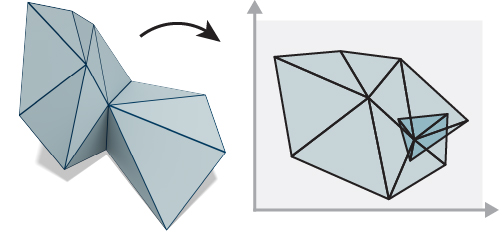}
	\label{fig:UVoverlap}
\end{wrapfigure}
Even if all the UV faces are oriented correctly, some of the faces may still overlap with each other depending on the flattening algorithm in use. We check whether the total angle sum of each interior vertex is $2π$ to determine the validity of a collapse.

\paragraph{Non-manifold edges}
\begin{wrapfigure}[5]{r}{1.33in}
	\raggedleft
    \vspace{-10pt}
	\hspace*{-0.7\columnsep}
	\includegraphics[width=1.45in, trim={6mm 0mm -1mm 0mm}]{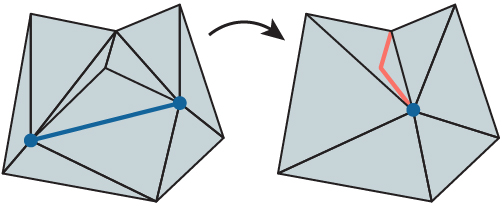}
	\label{fig:linkCondition}
\end{wrapfigure}
To prevent the appearance of non-manifold edges, \update{we must check the \emph{link condition} \cite{dey1999topology, HoppeDDMS93}.} Briefly, \update{the} link condition says that if an edge $e_{ij}$ connecting vertices $i,j$ is valid, the intersection between the vertex 1-ring of $i$ and the vertex 1-ring of $j$ must contain only two vertices, and the two vertices cannot be an edge.

\paragraph{Skinny triangles}
To prevent badly shaped triangles from causing numerical issues, we need to keep track of the triangle quality for each edge collapse. The quality of a triangle is measured by
\begin{align}
    Q_{ijk} = \frac{4\sqrt{3}\ A_{ijk}}{l_{ij}^2+l_{jk}^2+l_{ki}^2}
\end{align}
where $A$ is the area of the triangle and $l$ are the lengths of triangle edges. When $Q → 1$, it approaches an equilateral triangle; when $Q → 0$ , it approaches a skinny degenerated one. For each edge, we check $Q$ for all the neighboring faces in both UV and Euclidean domains after the collapse. By default, a valid edge requires $Q > 0.2$ for all neighboring triangles.

\section{Comparison to \cite{lee1998maps}}\label{app:compareMAPS}
One possible solution to construct a bijective map between the input and the decimated model is via \MAPS \cite{lee1998maps}. However, \MAPS constructs the parameterization via successively removing the maximum vertex independent sets. The main reason for removing the maximum independent set is to bound the number of levels of the mesh hierarchy, but it leads to limitations such as sensitivity to the input triangulation.

\begin{figure}
    \centering
    \includegraphics[width=3.33in]{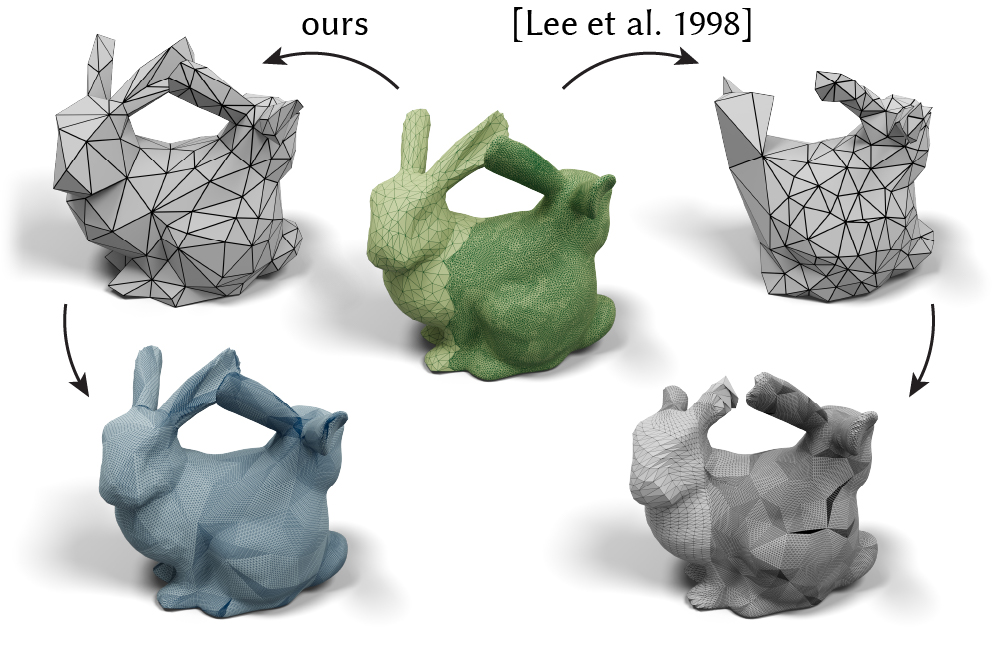}
    \caption{We decimate the mesh down to the same number of vertices and compare our method with \MAPS on the task of subdivision remeshing. Our method creates a more uniform parameterization (left), but \MAPS is more sensitive to the input triangulation (right).}
    \label{fig:uniformRemesh}
    \vspace{-5pt}
\end{figure}

One experiment to verify this is to apply subdivision remeshing presented in Sec. 4.1 in \cite{lee1998maps}.
In \reffig{uniformRemesh} we create a stress test using a very uneven triangulation, and \MAPS suffers from creating non-uniform parameterization.
In contrast our successive self-parameterization enjoys the benefits of area-weighted \qslim to obtain a more uniform parameterization.
% decimation \cite{garland1997surface} to obtain a more uniform parameterization.
%

\section{Data Generation from Random Collapses} \label{app:trainData}
The training data for neural subdivision is a sequence of subdivided meshes where the vertex positions are computed using successive self-parameterization (\reffig{trainData}). For each dense input mesh, we perform semi-random edge collapses in order to generate many different coarse meshes. The goal is to help the network to be robust to different discretizations. In \reffig{manyShapes} we show input meshes (left) can be decimated differently to get many coarse meshes that have different number of vertices and with different triangulations.

Our semi-random edge collapse starts by randomly selecting 100 edges and finding the one with the minimum \emph{quadric error} \cite{garland1997surface} to collapse. For each edge collapse, we insert the new vertex the same way as \qslim. We terminate the edge collapses when a randomly selected target number of vertices between 150 and 300 is reached.

\begin{figure}
    \centering
    \includegraphics[width=3.33in]{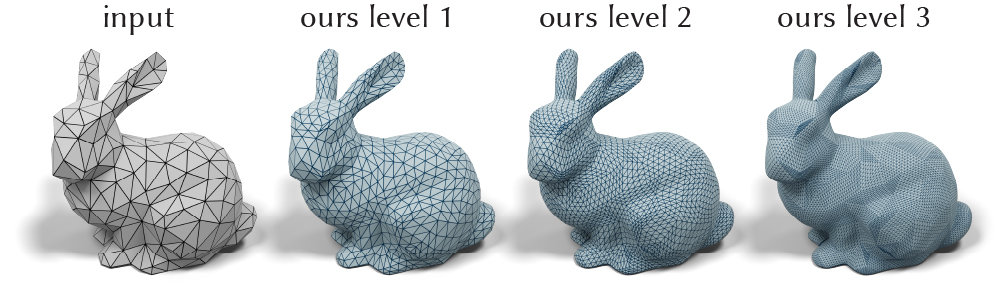}
    \caption{Although most experiments are trained on performing 2-level subdivisions, our neural subdivision network can still be trained on more level of the subdivisions. }
    \label{fig:level3}
    \vspace{-5pt}
\end{figure}
\section{Experimental Setup} \label{app:experiments}
Our experimental setup is consistent throughout the document. The training shape is presented in green in every figure. For each shape, we use the parameters described in \refapp{trainData} to generate 200 training discretizations and train for 700 epochs. Our method can learn to produce several subdivision levels \reffig{level3}, but we set the number of training subdivisions to two levels for consistency across the experiments. If the experiment consists of multiple training shapes, such as the experiments in \reffig{moreShapes} and \reftab{quantitative}, we evenly distribute the number of training discretizations so that they still sum up to 200 discretizations in total.

\section{Learning Classic Loop Subdivision} \label{app:Loop}
\begin{figure}
    \centering
    \includegraphics[width=3.33in]{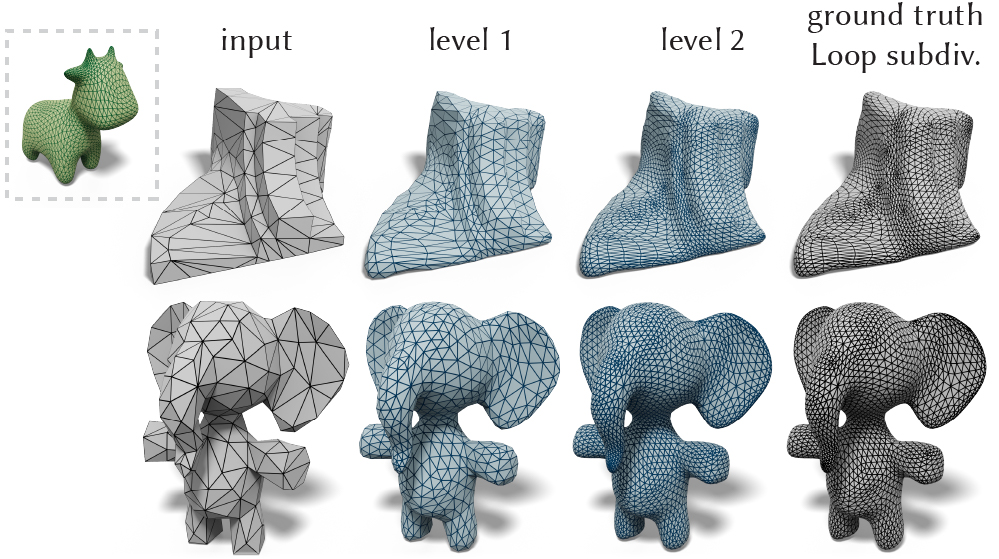}
    \caption{\update{When trained on meshes created by classic Loop subdivision (green), our network can reproduce the Loop scheme on new meshes, and creates visually indistinguishable results (blue) compared to the ground truth created by the classic Loop method (right).}}
    \label{fig:learnLoop}
    \vspace{-5pt}
\end{figure}

\update{Although we have shown in \refsec{results} that neural subdivision is able to subdivide a mesh adaptively, one might be interested in seeing whether neural subdivision can also learn to reproduce classic Loop subdivision with appropriate training data. In \reffig{learnLoop}, we trained our network on a sequence of meshes created with Loop subdivision. 
Given an original mesh, we create 200 mashes using random edge collapses, then subdivide each coarsened mesh for two levels using Loop to obtain the corresponding ground truth subdivided sequences for measuring the reconstruction loss.
We see that when testing on novel meshes, the network is able to reproduce the Loop scheme to create visually indistinguishable results. The average per-vertex numerical error is just $0.3\%$ of the bounding box diagonal.}

\vspace{25pt}
\section{Ablation Studies (continued)} \label{app:ablation}
\begin{wrapfigure}[7]{r}{1.33in}
	\raggedleft
    \vspace{-12pt}
	\hspace*{-0.7\columnsep}
	\includegraphics[width=1.45in, trim={6mm 0mm -1mm 0mm}]{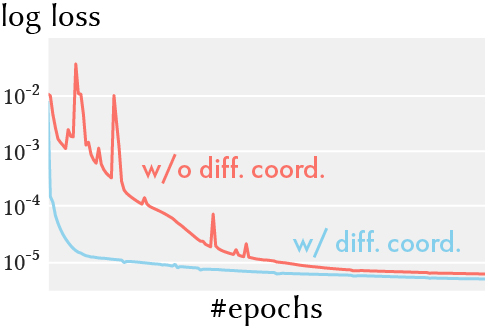}
\end{wrapfigure}
This section summarizes the ablation studies of other design decisions we made in the network design. These components are not as crucial as the components mentioned in the main text, but they still offer improvements while training. The first analysis is the influence of differential coordinates in the input (see \reffig{flapInput}). Our result in the inset indicates that adding differential coordinates can improve convergence.

\begin{wrapfigure}[7]{r}{1.33in}
	\raggedleft
    \vspace{-12pt}
	\hspace*{-0.7\columnsep}
	\includegraphics[width=1.45in, trim={6mm 0mm -1mm 0mm}]{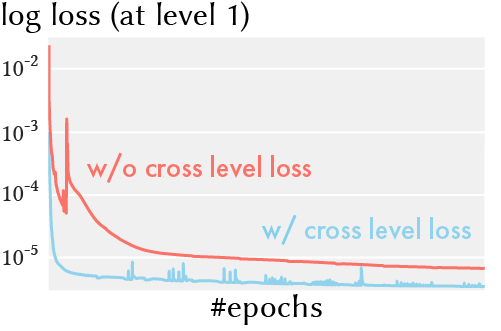}
\end{wrapfigure} 
We also measure the effect of adding cross-level loss compared to only measuring the loss at the final level. In the inset, we visualize the error in the intermediate level. The result suggests that adding cross-level loss can improve subdivision results in the intermediate levels, which is important for creating meshes with different levels of detail (see \reffig{allLevels}).

\begin{wrapfigure}[7]{r}{1.33in}
	\raggedleft
    \vspace{-12pt}
	\hspace*{-0.7\columnsep}
	\includegraphics[width=1.45in, trim={6mm 0mm -1mm 0mm}]{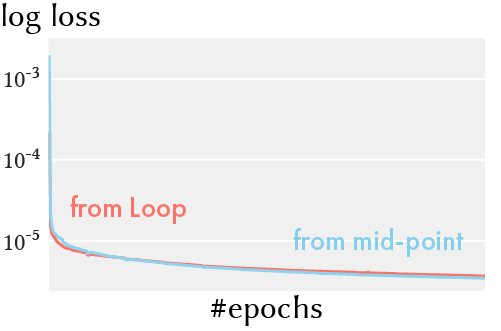}
\end{wrapfigure}
The third study is on the starting position of the predicted displacement vector as shown in \reffig{flapOutput}. Specifically, we compare predicting the displacement from the mid-point of an edge with predicting displacement from the Loop-subdivided mesh. Our result in the inset suggests that using different starting positions has no influence to the quality of the output. Thus we choose the mid-point for simplicity.

\begin{wrapfigure}[7]{r}{1.33in}
	\raggedleft
    \vspace{-12pt}
	\hspace*{-0.7\columnsep}
	\includegraphics[width=1.45in, trim={6mm 0mm -1mm 0mm}]{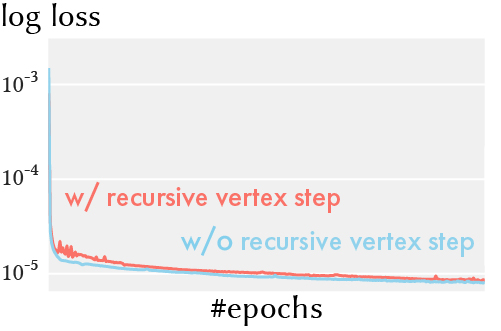}
\end{wrapfigure}
The fourth study is on the number of vertex steps to perform. In \reffig{pipeline}, we can actually recursively perform the vertex step to gather information from larger rings. However our experiments in the inset indicates that recursively performing the vertex step does not offer improvements. Thus we only perform the vertex step once. We suspect that the 2-ring information on the coarse mesh (one from initialization, one from the vertex step) may already be sufficient for the network to perform subdivisions.

\begin{wrapfigure}[7]{r}{1.33in}
	\raggedleft
    \vspace{-12pt}
	\hspace*{-0.7\columnsep}
	\includegraphics[width=1.45in, trim={6mm 0mm -1mm 0mm}]{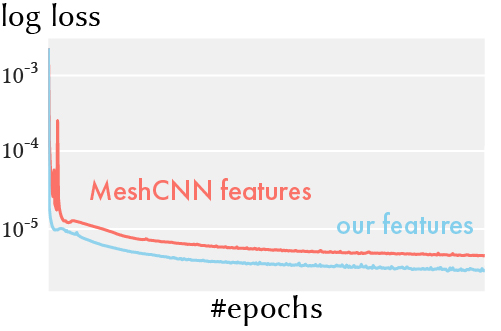}
\end{wrapfigure}
In MeshCNN, \citet{hanocka2019meshcnn} propose a set of features to characterize an undirected edge (via features of a flap), including the dihedral angle, two inner angles, and two edge length ratios (see Sec. 3 in \cite{hanocka2019meshcnn}). We tried their proposed features in our neural subdivision network. In the inset, we observe that using our features, edge vectors and the vectors of differential coordinates, converges to a better solution. 

\end{document}